# Public vs Private Bodies: Who Should Run Advanced AI Evaluations and Audits? A Three-Step Logic Based on Case Studies of High-Risk Industries

**Merlin Stein[1]*, Milan Gandhi[1], Theresa Kriecherbauer[1], Amin Oueslati[2] and Robert Trager[1]**

[1]University of Oxford [2]Hertie School [*]Primary author, merlin.stein@bsg.ox.ac.uk. AIES 2024.

## Abstract

Artificial Intelligence (AI) Safety Institutes and governments worldwide are deciding whether they evaluate and audit advanced AI themselves, support a private auditor ecosystem or do both.

Auditing regimes have been established in a wide range of industry contexts to monitor and evaluate firms' compliance with regulation. Auditing is a necessary governance tool to understand and manage the risks of a technology. This paper draws from nine such regimes to inform (i) who should audit which parts of advanced AI; and (ii) how much capacity public bodies may need to audit advanced AI effectively.

First, the effective responsibility distribution between public and private auditors depends heavily on specific industry and audit conditions. On the basis of advanced AI's risk profile, the sensitivity of information involved in the auditing process, and the high costs of verifying safety and benefit claims of AI Labs, we recommend that public bodies become directly involved in safety critical, especially gray- and white-box, AI model evaluations. Governance and security audits, which are well-established in other industry contexts, as well as black-box model evaluations, may be more efficiently provided by a private market of evaluators and auditors under public oversight.

Secondly, to effectively fulfill their role in advanced AI audits, public bodies need extensive access to models and facilities. Public bodies' capacity should scale with the industry's risk level, size and market concentration, potentially requiring 100s of employees for auditing in large jurisdictions like the EU or US, like in nuclear safety and life sciences.

**Figure-based abstract: A three-step logic and its result for advanced AI audits/evaluations.**

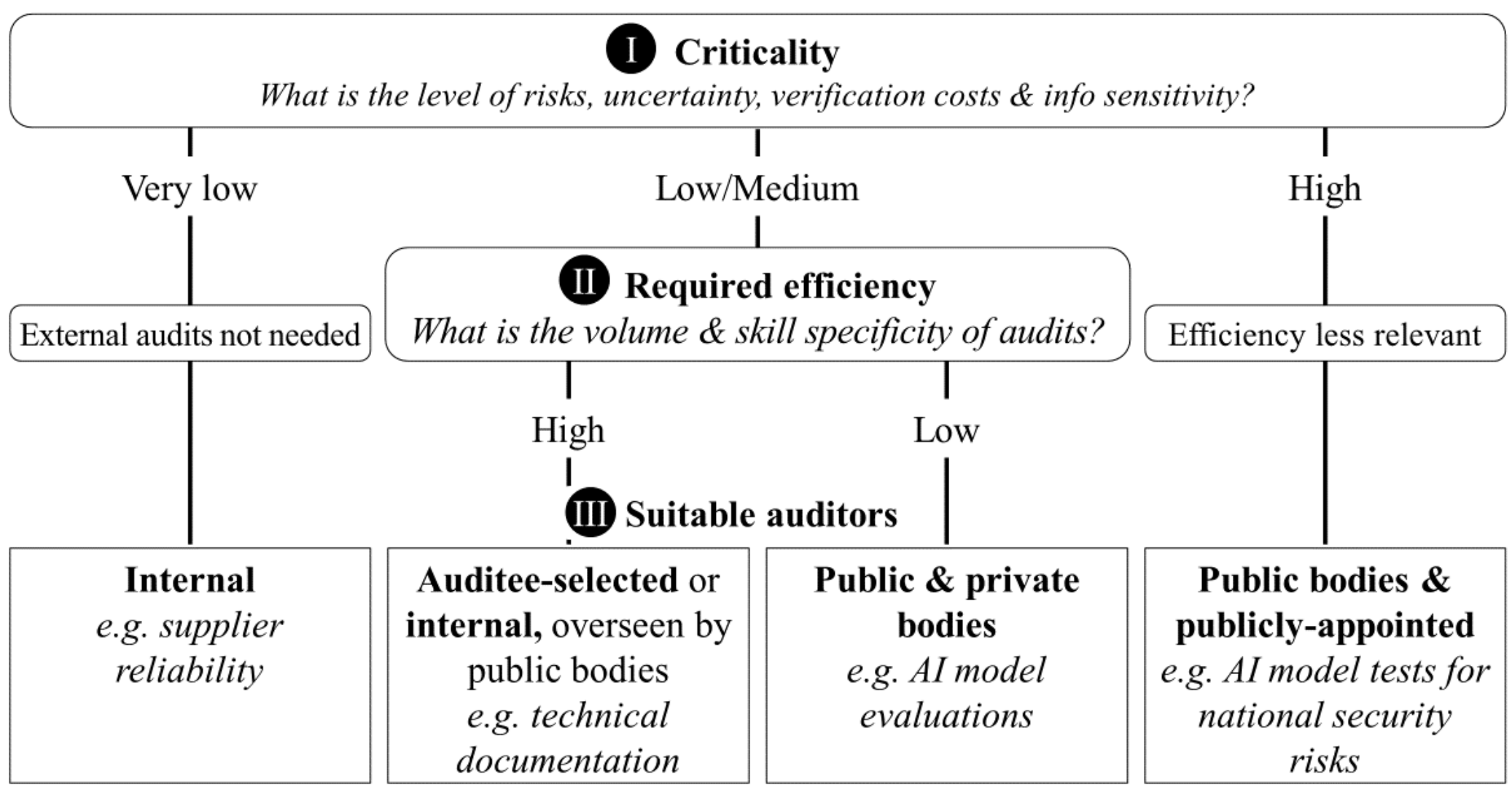

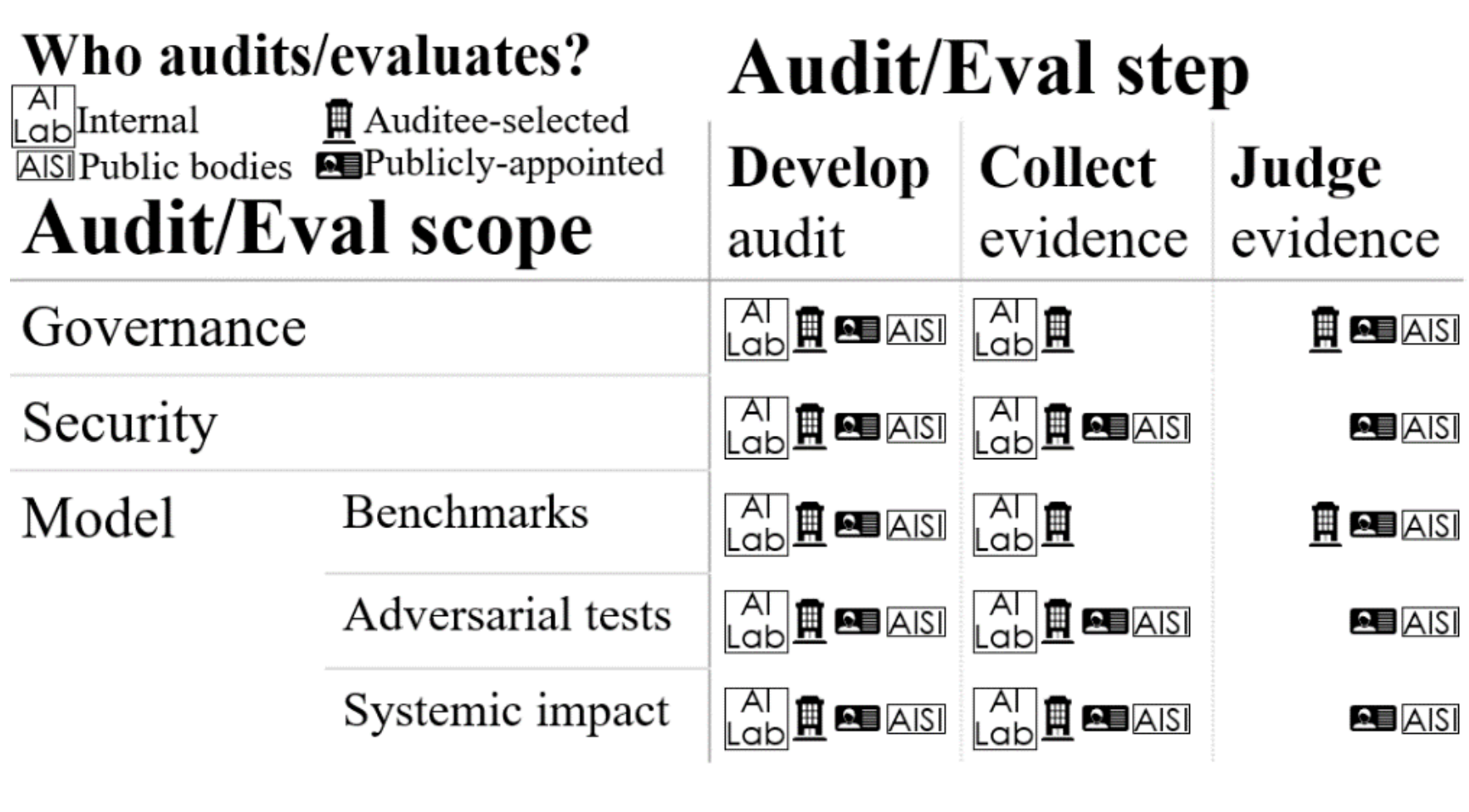

# Executive Summary

## Why is it important who audits advanced AI?

Governments worldwide are deciding how to govern advanced artificial intelligence (AI). Establishing an AI auditing regime is one important policy lever to assess conformity. Governments must decide what party, public or private, is best placed to carry out AI audits (Shevlane et al. 2023, Stephenson 2011). The question of who audits is important because auditing regimes can differ significantly in their scope, efficiency and propensity to deliver safety and benefits (Raji et al. 2022, Kleinman, Lin and Palmon 2014).

## What are the audit steps for advanced AI?

We broadly define advanced AI auditing to include ‘evaluations’, ‘scrutiny’, ‘inspections’, and ‘monitoring’. We segment AI audits into those that focus on: (1) governance; (2) security; and (3) the AI model. Furthermore, we suggest that there are three main steps in any kind of AI audit: (a) developing the audit; (b) collecting evidence; and (c) judging evidence. The central challenge throughout the steps is: Efficiently building and using resources to make sure the *right* audits are run *well*.

## Who can run advanced AI audits?

Auditors can be the AI developer internally, externals selected or paid by the AI developer, and externals independent of the AI developer such as, for example, public bodies, publicly-appointed auditors or independent civil society and academic actors.

| | Auditor characteristics | | | |
|---|---|---|---|---|
| **Auditor type** | **Independ.** | **Resources** | **Competence** | **Access** |
| **Public bodies** | Public scrutiny | Inflexible | Built if high salience | Clearances, mandates |
| **Publicly-appointed** | Quality for re-selection | Inflexible tendering | Specialized experts | Depends on security clearance |
| **Auditee-selected** | Lenient for re-selection | Flexible ecosystem | | |
| **Internal** | Private interests | Fast deployment | Product-specific | Internal access |

*Level of auditor characteristics: High Medium Low*

Figure A: Typical auditor characteristics suggested by the auditing literature. Auditors can build their capacity, competence and access

## Who should run which advanced AI audits?

We conducted nine case studies of high-risk auditing regimes in the United States, to qualitatively identify who audits under what conditions. We do not measure audit effectiveness directly, but the alignment of demand-side conditions and supply side auditor characteristics that underly effectiveness. As Figure B illustrates, criticality and market concentration determine involvement of public bodies. For advanced AI, these factors might be different for different audits. We suggest that an effective oversight model combines auditing by a public body with auditing by private firms. Auditing could involve a pool of auditors with firms being appointed to audits according to their expertise. We propose a three-step decision logic with suitable auditors illustrated in Figure D.

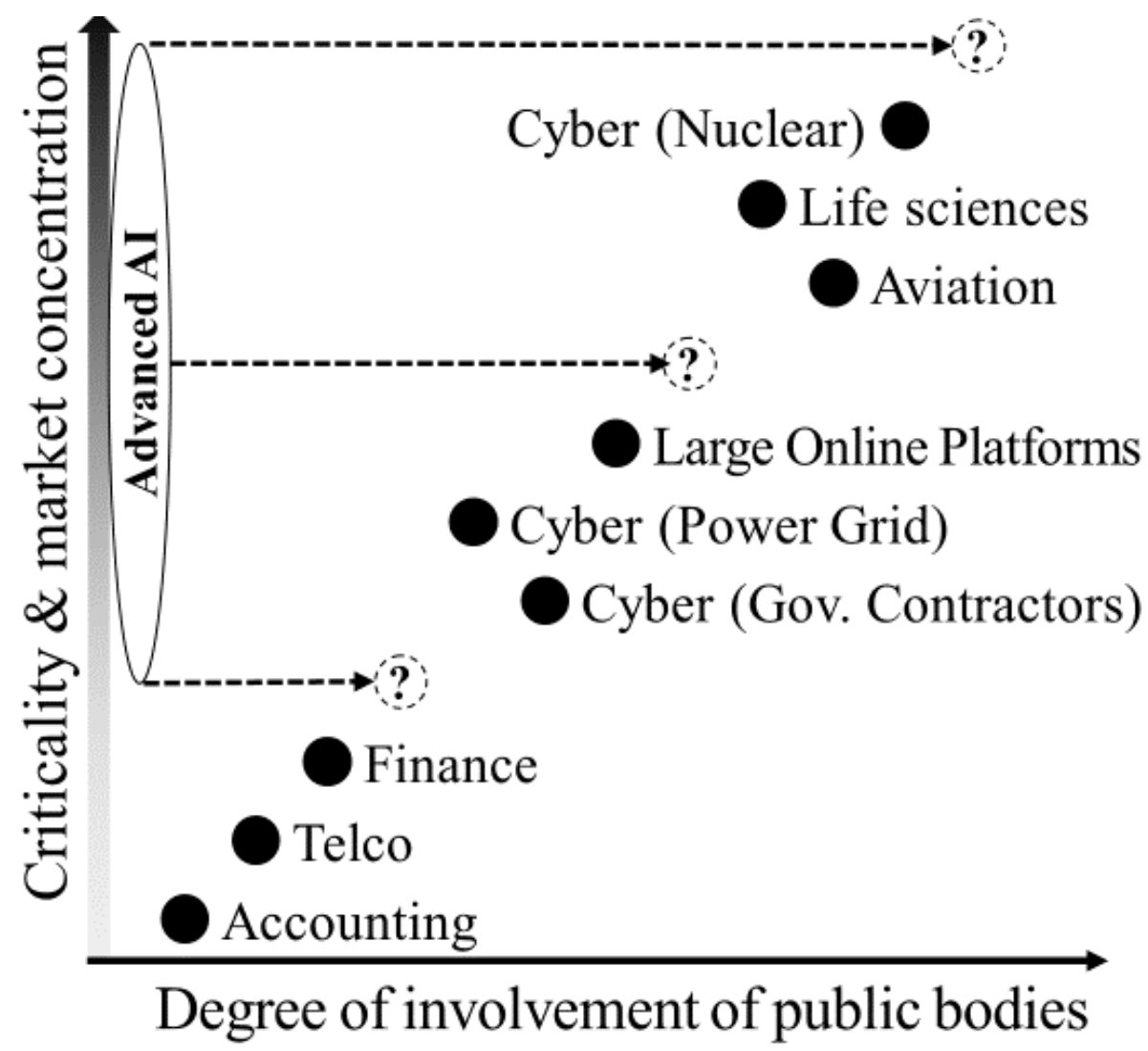

Figure B: Criticality (harm to externals, risk uncertainty, verification costs and info sensitivity) and market concentration is associated with high involvement of public bodies in auditing. Based on a quantified ranking.

**Step 1 - Criticality: How severe and uncertain are risks, and how much sensitive information and resources are required to verify developer claims?**

Across the nine auditing regimes studied, we observe that public bodies are more involved in developing audits, collecting evidence and judging evidence when criticality is high, i.e.: i) Audit information is sensitive; ii) It is costly to verify conformity with rules iii) Risk uncertainty is high, and iv) Potential harms to third-parties are high.

For example, in aviation, public bodies are heavily involved in all three steps in the auditing process: developing the audit; collecting evidence; and judging evidence. By contrast, in telco, radio frequency device security information is collected and judged by private auditors. Public bodies only judge private auditors.

What could explain these differences? The higher the criticality, the greater the need for independent auditing, i.e. by a public body or publicly-appointed auditors.

**Step 2 - Efficiency: Who has or can build the required capacity, competence and access?**

The nine case studies suggest that when criticality is low, efficiency becomes relatively more important. Public or private bodies with existing resources - especially capacity (staff and financial resources), competence (staff expertise) and access (to information) - can conduct audits more efficiently. Public and private bodies can actively build these resources, especially in new, salient fields - like advanced AI.

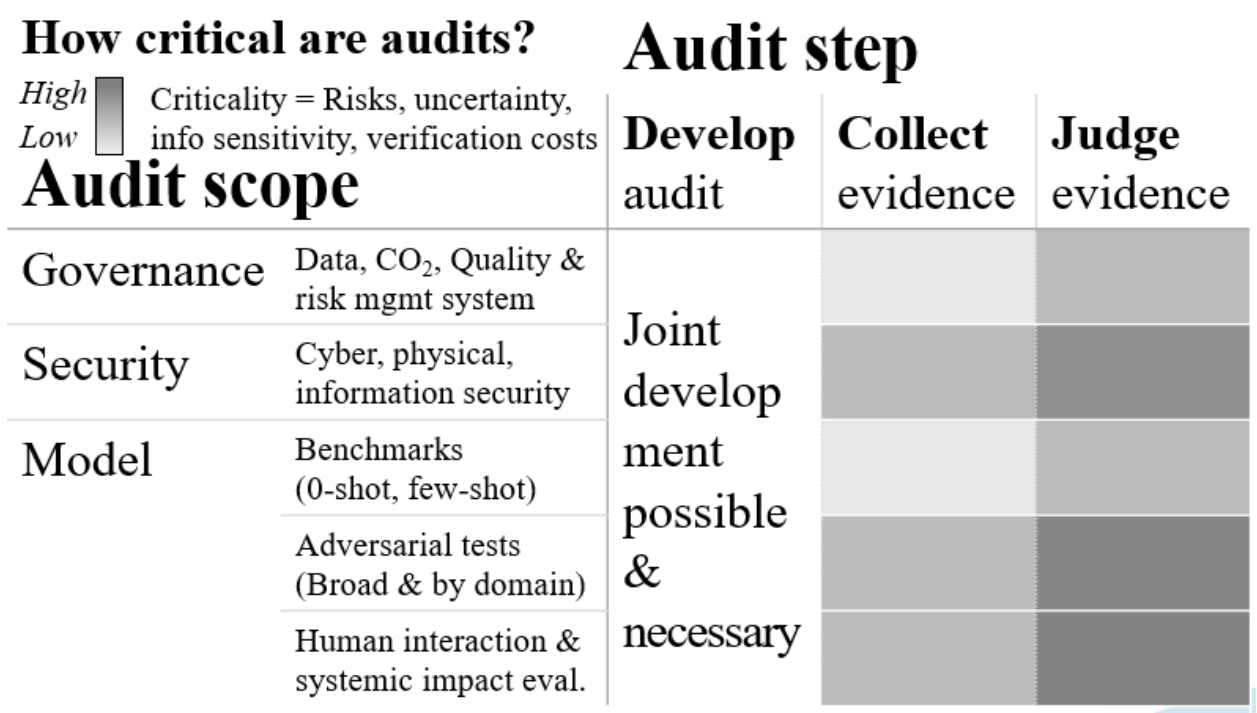

Figure C: Criticality for each audit step and scope. Risks, uncertainty, info sensitivity and verifications cost influence criticality.

**Step 3 – Suitable auditors**

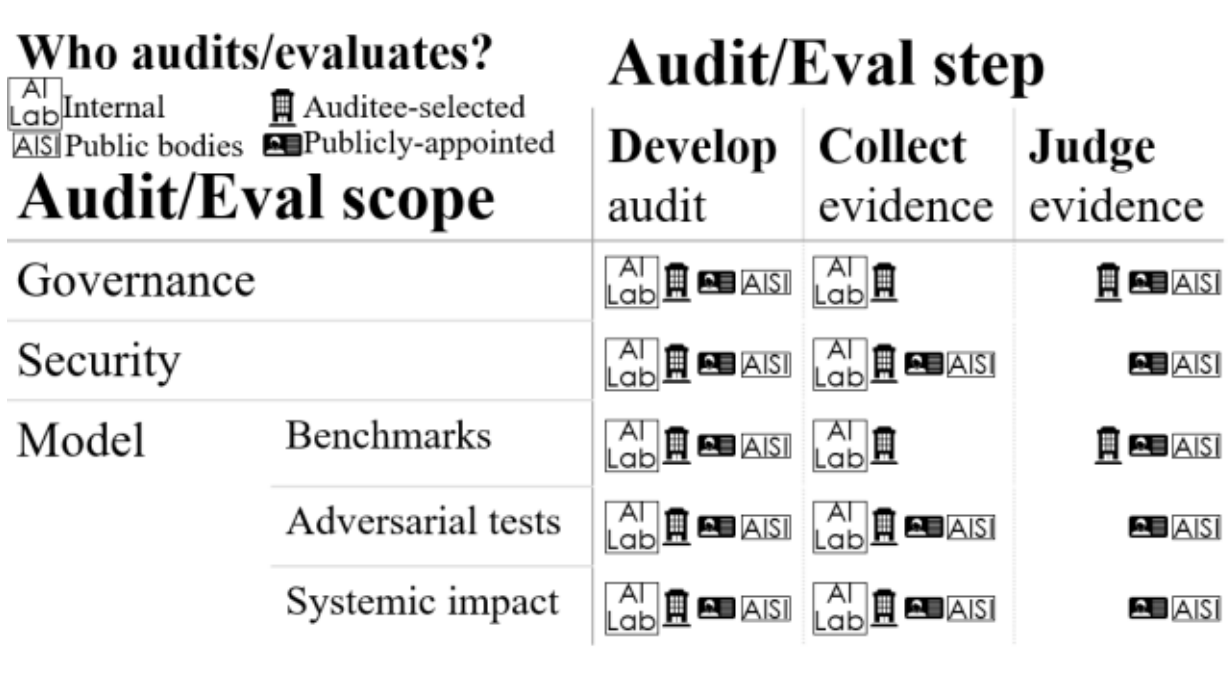

Figure D: Suggested involvement of auditors by type.

## Recommendations

The required capacity, competence and access a public body needs to establish an effective AI auditing regime depends on its beliefs about future AI industry development. If risks and risk uncertainty remain high, and the number of developers and new advanced AI models are low, then public bodies may need to be more involved in auditing. This means they must build more capacity, competence and access. When the volume of audits is high, support by private auditors is required.

**Recommendations for AI Safety Institutes**

- **Prioritize critical audits.** Focus on direct auditing of AI systems where risks, uncertainties, info sensitivity and verification costs are high, like in nuclear or aviation.
- **Build internal capacity & competence.** To reduce the uncertainty of advanced AI and develop a science of evaluations, build expertise and auditing capacity by doing audits of advanced AI.
- **Ensure and use access** to auditees' and auditors' information to judge evidence and audits
- **Build the ecosystem** via standards like "Inspect", training programs, partner programs and indirect access
- **Review audits** conducted by various players, to review auditors practices and quality and point to best-practices

**Recommendations for advanced AI regulators**

- **Guarantee access.** Provide access to researchers and public bodies to preempt regulatory demands
- **Consider statutory auditing.** Connect audit results to post-audit transparency and regulatory enforcement
- **Fund AISIs and externals auditors** via a mix of public budget and industry fees

**Recommendations for AI auditors/evaluators**

- **Build specific competence.** Focus on 1-2 subfields of auditing, to drive the maturity in the nascent field.
- **Remain as independent as possible.** Avoid conflicts of interest and be transparent on clients and funding.
- **Develop audits using open models and share methods.** While access restrictions limit testing some models, audits and evals can be developed for open source models and transferred to closed source models.

**Recommendations for advanced AI Labs**

- **Share access and expertise.** Product-specific experience shared on blogs, developer forums, …
- **Share access.** To develop a set of trusted auditors, share access in stages
- **Commit to post-audit actions.** Define **r**esponsible scaling policies pre-audit for specific audit results along quantified scales with predefined post-audit actio

# 1 Introduction

Governments across the world are exploring new regulations to mitigate the risks of advanced artificial intelligence (Weidinger et al. 2022, UK Government 2023). Establishing an auditing regime is one tool available to policymakers to facilitate and enforce firms' compliance with AI rules. For the purposes of this paper, we define an 'AI auditing regime' as the institutional framework by which advanced AI developers and providers ("AI Labs"), particularly their AI models, are subjected to evaluation by externals. Under this definition, there are numerous design choices available to policymakers (Birhane et al. 2024). We explore two sets of choices, (1) Who should audit which parts of advanced AI? (2) What resources, competence and access should the public body develop to carry out its role in auditing? We use 'public body' to refer to the government institution that is primarily responsible for auditing, whether through rule-setting and oversight or undertaking audits. *We use the term "audit" to include exploratory evaluations, targeted auditing and monitoring.* Our contributions are:

- **Auditing Regime Case Analysis and Design Factors:** We analyze nine industry cases to identify dimensions along which auditing regimes differ, and quantify industry and audit factors explaining the differences. These are an extension of hybrid governance theory.
- **Three-Step Logic for Auditing Regime Design:** We propose a three-step logic to determine who is best placed to audit depending on the industry context, demand for auditing and the type of auditing required. We apply this logic to derive policy recommendations for designing advanced AI auditing regimes.
- **Estimate of Required Capacity in AI Safety Institutes or Other Public Bodies for Advanced AI:** We empirically estimate the resource, competence and access requirements for a public body in an advanced AI auditing regime.

This paper is structured as follows:

- Section 2 locates the study in the literature
- Section 3 explains the methodology and limitations
- Section 4 proposes demand-side and supply-side factors determining who could and should audit
- Section 5 explores nine high-risk auditing regimes and extrapolates a three-step logic on who should audit
- Section 6 applies the three-step logic to advanced AI
- Section 7 outlines resource, competence and access requirements for public bodies in advanced AI auditing
- Section 8 describes open questions and concludes

# 2 Related Literature

We use the term 'advanced AI' to refer to state-of-the-art general-purpose AI models, aligning with the definition of the International Scientific Report on the Safety of Advanced AI (DSIT 2024). As this report and other research analyzes, firms that develop advanced AI risk the imposition of unpredictable and potentially severe costs on unconsenting third parties (DSIT 2024). Such externalities require government intervention (Pigou 1920). Embedded in a spectrum of measures (Gunningham, Grabosky and Sinclair 1998), one important intervention is AI auditing (Costanza-Chock et al. 2023). However, advanced AI is one of the fastest evolving and complex general-purpose technologies. Its externalities are difficult to reliably estimate (DSIT 2024, Hobbhahn and Scheurer 2024). Thus, advanced AI auditing regimes need to address the challenge of running the *right* audits *well* under resource constraints ("audit effectiveness")[1].

**Running the right audits.** Sufficient and flexible capacity is necessary to keep up with the speed of AI progress and therefore an expanding list of dangerous capabilities and downstream sociotechnical risks (EpochAI 2023). Auditors need to be competent and have access to assess capabilities and risks. Running the right audits means reducing uncertainties, e.g. through standardization.

**Running audits well: Independence vs. resource efficiency.** The most independent auditors aligned with public interest – public bodies, publicly-appointed auditors, academics or civil society – may be less efficient and flexible than private auditors. However, private auditors fail to produce high quality auditing when they share conflicts of interest with auditees (DeFond 2010). Thus, balancing independence and efficiency can mean trading-off audit quality and efficiency. This trade-off shapes auditing regimes. Audit quality and independence is more important for audit steps that are critical for public safety (Brundage et al. 2020, Power 1999). The industry setting and auditing ecosystem, including the distribution of resources and skills may influence this trade-off (Power 1999).

[1] An effective audit requires accurately assessing relevant benefits and harms ("audit quality"), while minimizing costs and delay ("audit efficiency").

Previous literature observes significant variability in the design and effectiveness of auditing regimes across industry contexts (Kleinman, Lin, and Palmon 2014; Raji et al. 2022). Key factors influencing auditing effectiveness include auditor independence (Duflo et al. 2013, Short et al. 2016), resources (Anderljung et al. 2023), competence (DeFond and Zhang 2014) and the auditor's access to evidence for the audit (Lamoreaux 2016, Raji et al. 2022), and the auditor's access to the evidence required for audits (Simnett, Carson, and Vanstraelen 2016; Simnett and Trotman 2018; Hansen, Kumar, and Sullivan 2008).

Questions about the public body's optimal role in an advanced AI auditing regime remain under-explored (Hadfield and Clark 2023). Extant auditing literature emphasizes auditor characteristics like independence in explaining regime effectiveness. Our research examines underlying characteristics of the industry and audit, exploring their implications for auditing regime design.

For this purpose, we connect with the hybrid governance literature and new institutional economics. Effective governance is only partly determined by the characteristics of the oversight or auditing body, mainly by the alignment of these characteristics with the underlying conditions of transaction costs and asset specificity (Menard 2004, Quélin et al. 2019). As hybridity shapes AI governance (Radu 2021) and auditing too (Rajala and Kokko 2021), we adapt Menard's (2004) hybrid governance framework for the auditing context.

# 3 Methodology, Scope and Limitations

To analyze which advanced AI audits should be performed by public and which by private bodies and its resource implications, we surveyed examples of auditing regimes across nine different industries, focusing our analysis on critical infrastructure sectors in the United States ("US")[2]. This comparative case study approach has proven effective for similar prescriptive questions on regulatory regimes (Levi-Faur 2003, Hill and Varone 2021). Given the small number of high-risk regimes and difficulty in capturing nuances in their variations quantitatively, we deploy an exploratory, inductive mixed-method approach. Based on existing literature, case studies and in line with hybrid governance theory, we identify demand-side factors determining auditing responsibilities across and within industries. To understand variations at a high level, we quantify the demand-side factors for each case, and observe their association with the degree of public body involvement in auditing. To explain this link and derive more granular implications for advanced AI auditing, we qualitatively analyze auditing supply and estimate public bodies' capacity requirements.

Our case study research describes what *is* the case across contexts, and does not measure effectiveness of audit regimes directly, nor establish a causal link quantitatively. Instead, we follow Menard (2004) and assume that effective governance is largely determined by the alignment of the characteristics of the auditing body with underlying demand-side factors. Further, our capacity estimates are initial, rough approximations, and require more dedicated research.[3] Appendix B further details the methodology and limitations.

[2] We focus on US oversight regimes, given the country's leading AI development capacity (Tortoise 2023). As part of the case studies, we briefly compared each industry's US regimes to their counterparts in the EU and UK, finding no major deviations, even though regimes in the US are slightly more liberal in most industries (e.g., OECD 2014). We add audits of online platforms in the EU as an additional case that is less present in the US.

The US, for example, categorizes AI as a critical and strategically important technology (NSTC 2024). Given Advanced AI's potential risks, we focus on sectors classified as critical infrastructure in the US, EU and UK. Of these, we use sectors with especially high speed of innovation based on patents filed (Marco et al. 2017). Within each sector, we pick a typical product or security system for clarity. This leads to our case studies on transport (airworthiness certification of civil airplanes), communications (authorization of radio frequency devices in telecommunications), IT (cyber security infrastructure - for nuclear power plants, government contractors and bulk power systems; and large online platforms), financial services (audits of public companies' annual reports, risk monitoring of financial securities products), and life sciences (regulatory approvals process for medical software). We focus on audits in civilian contexts, within a single jurisdiction, aimed at general-purpose models. We assume that advanced AI will be developed by private entities.

[3] Our quantification of capacity requirements of public bodies in section 7 is only a first, simplistic estimate. We acknowledge that this approach is imperfect as, e.g., the size of the AI industry does not necessarily correlate with the demand for AI audits, and depends on the jurisdiction.

# 4 Framework of Analysis: Auditing Factors

We propose that differences across regimes on *who audits* and *with which resources* can be understood by considering the nature of risks in an industry being addressed by the audit and challenges inherent to auditing (demand-side factors); and the characteristics of auditors (supply-side factors). We analyze them in each case study.

## 4.1 Demand-Side Factors: The Nature of the Risk

We suggest that the demand for and emergence of an auditing regime is shaped by an industry's risk profile and its perceived importance by the public (Ramanna 2015). In addition, the industry market size and concentration may impact the volume of audits demanded.

An adjacent set of demand-side factors are inherent to a specific audit. These relate to the availability of auditors and the skills they require to conduct audits, which vary according to the complexity of auditing methods. Furthermore, there is a perennial information problem – to conduct the audit, the auditor must obtain information in the control of the auditee and verify the auditee's claims. Finally, collecting, managing and, in some regimes, publishing this information may pose its own set of risks if the information is sensitive to, for example, intellectual property and national security concerns.

We posit that demand-side factors influence the trade-off between audit quality and audit efficiency. High levels of risk uncertainty, potential externalities, verification costs, and information sensitivity necessitate prioritizing audit quality, which is achieved through auditors' independence, competence, and access. Conversely, in large, less concentrated markets, audit efficiency becomes paramount, achieved through auditors' existing and adaptable capacity and relevant skills. These requirements for auditor characteristics subsequently dictate the allocation of audit responsibilities and the resources public bodies may need to develop.

| **Demand-side factors** | |
|---|---|
| ***Industry conditions*** | |
| Risk uncertainty | Predictability and clarity regarding risks and risk measures *(ISO standard length and share of standards under development)* |
| Potential for externalities | Risk severity & third-party exposure to harm when risks materialize *(National Risk Register: Impact and likelihood of risk)* |
| Public salience | Level of importance the public places on an industry's risks *(# search results on Google News across the last 5 years)* |
| Market size / concentration | Distribution of and total industry revenue across firms *(Herfindahl Index)* |
| ***Audit conditions*** | |
| Verification costs | Cost of establishing an auditee's conformity with rules *(Invasiveness of audit procedure)* |
| Information sensitivity | Potential harm from unauthorized use of information required for audit *(Governmental document sensitivity classifications)* |
| Skill specificity | Rarity and level of specialized expertise required for audit *(Level of market-based salary)* |

Table 1: Definition and quantification proxy (in brackets) of demand-side factors. Each proxy value is categorized into high, medium and low for simplicity. Criticality refers to the first four factors.[4]

## 4.2 Supply-Side Factors: Auditor Characteristics, Archetypes and Auditing Responsibilities

The factors outlined above define the demand for and challenges of auditing within an industry context. An appropriate auditing regime fulfills this demand by incentivizing independence and sufficient capacity (resources, competence and access) of auditors.

| **Supply-side factors (Audit characteristics)** | |
|---|---|
| Independence | Absence of conflicts of interest (e.g. due to selection/payment by auditee), in public interest |
| Resources | Auditor's human, financial, computational and other resources; and flexibility of using resources |
| Competence | Auditor's skills and experiences in the kinds of audits demanded |
| Access | Extent of the auditor's access to evidence required for the audit (e.g., to data, tech, offices, staff) |

Table 2: Definition of supply-side factors

[4] The demand-side factors are in line with Menard's three hybrid governance factors (2004), adapted for auditing. 1) Uncertainty is captured by risk uncertainty relating to the validity and reliability of information about risks. 2) Transaction costs are described on the extensive margin as the reasons for auditing transactions ("risk externalities"), on the intensive margin as the difficulty of the auditing transaction ("verification costs" and "information sensitivity"). 3) Asset specificity describes how the competence of auditors generalizes ("skill specificity"). However, economic hybrid governance theory is limited in focusing on economic hybridity and agents. This allows for this paper's actor-focused approach, but falls short of analyzing auditing from a within-organizational perspective (Bol et al. 2019) and power-distribution perspective (Levi-Faur 2011). To bridge the latter limitation, we build on regulation theorists (Behr 1985 and Stigler 1971) to establish factors which pertain to the existing power distribution, from a societal and an economic perspective. Following auditing scholars (Ramanna 2015), we separate power distribution into "public salience" and "market concentration and size".

We cluster different sets of auditor characteristics into four idealized auditor archetypes. In practice, a classification of public-appointed auditors as "highly independent" should be interpreted as a potential degree of independence, while currently many publicly-appointed auditors have conflicts of interest, e.g. due to simultaneous consulting work. Appendix C details our assumptions on auditor characteristics in depth.

| Auditor type | Auditor characteristics: Independ. | Resources | Competence | Access |
|---|---|---|---|---|
| **Public bodies** | Public scrutiny | Inflexible | Built if salient | Clearances, mandates |
| **Publicly-appointed** | Quality for re-selection | Inflexible tendering | Specialized experts | Depends on security clearance |
| **Auditee-selected** | Lenient for re-selection | Flexible ecosystem | Specialized experts | Depends on security clearance |
| **Internal** | Private interests | Directly available | Product-specific | Internal access |

*Level of auditor characteristics: High Medium Low*

Table 3: Auditor archetypes[5] and potential, idealized characteristics suggested by the auditing literature. Auditors can build and change their characteristics (Details in Appendix C).

**Auditing Responsibilities Along the Audit Lifecycle**

We suggest that the lifecycle of all audit processes involves the following three stages (Raji et al. 2020, Ojewale et al. 2024)[6][7]:

1. **Developing** auditing methods and rules
2. **Collecting** evidence ('auditable artifacts') for the audit in accordance with the auditing method
3. **Judging** the evidence, producing an audit report.

The combination of audit scope (for advanced AI models: governance, security and model - see below) and audit lifecycle defines the auditing responsibility space. Different auditor archetypes can fulfill each responsibility. In the following, we observe who fulfills different responsibilities across case studies.

# 5 Auditing Regime Case Study Findings

## 5.1 Comparative Case Study Findings

The framework above is applied to each case, as illustrated below for one example case. In addition, each case is qualitatively examined along its historical emergence, responsibility setup and audit effectiveness. There are many factors shaping an auditing regime, like the degree of information access or continuity of audits (See Appendix A.1 for details on each case).

**Case example: Cybersecurity audits in nuclear energy**

| | | |
|---|---|---|
| Demand-side factors | Risk uncertainty | Medium *(<50% of ISO standards under development but >2000 pages)* |
| | Potential for external. | High *("catastrophic" classification)* |
| | Verification costs | Medium *(Inspections and simulations)* |
| | Information sensitivity | High *(Classification "restricted")* |
| | Market concentration | High *(Herfindahl Index of 1500)* |
| | Skill specificity | Medium *($122k salary for a nuclear cybersecurity analyst)* |
| | Public salience | High *(43M news search results)* |
| Supply-side | High criticality, thus independence important. High market concentration, thus inflexible capacity okay. High salience allows for capacity build-up in public bodies. | |
| Auditor | Who judges audit | Public bodies |
| | Who collects evidence | Public bodies & Internal |
| | Who develops audit | All |
| | Who audits the auditor | Public bodies |

Table 4. Case example: Cybersecurity in nuclear energy.

[5] Given our focus on regulatory-demanded, statutory audits, we do not specifically list civil society or academic auditors - though the category of "publicly-appointed auditors" could be expanded to include them, while results remain similar.

[6] The demands of each stage depend on the type of audit undertaken and its purpose. Consider, for example, an AI model audit that utilizes benchmarking. Firstly, the auditor must select or develop the framework of benchmarks, resulting in a dedicated software package. This undertaking is technical and conceptual, requiring a match between the purpose of the audit and the metrics adopted. Multiple kinds of subject matter expertise may be required, e.g., relating to the model, auditing method and domain of interest for the audit (such as CBRN risks). Secondly, the auditor runs the AI model through the selected benchmarks to gather performance data. This may require substantial engineering effort, from preparing and formatting benchmark datasets through to ensuring benchmarking tools interface with the AI model. Thirdly, the auditor judges the performance data, decides on the need for additional tests and corrective measures, and produces the audit report.

[7] We exclude post-audit actions like transparency and enforcement considerations from this analysis for reasons of brevity. An audit of the auditor follows similar steps.

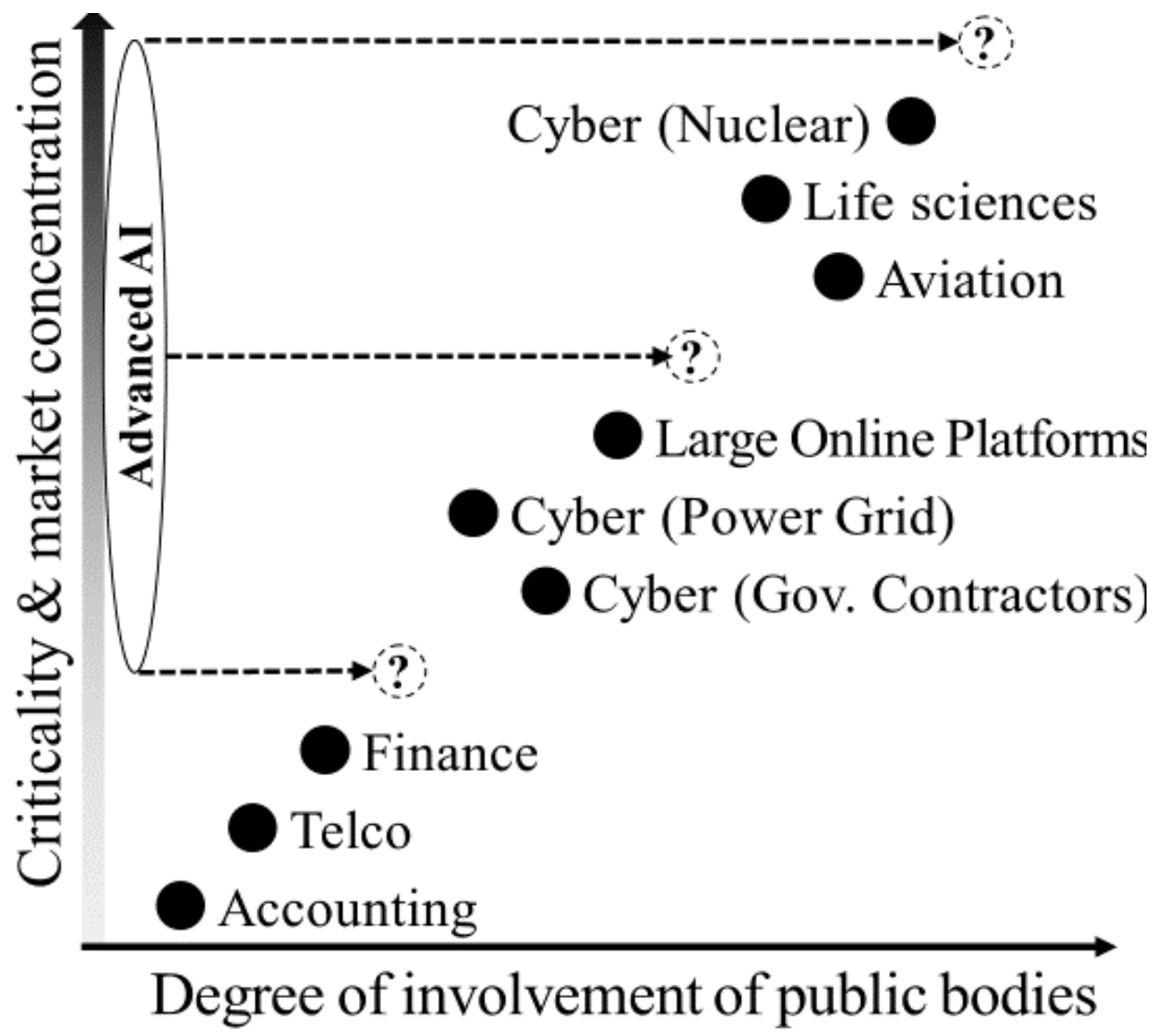

Figure 1: High criticality (risk externality, risk uncertainty, verification costs and info sensitivity) and market concentration of an industry is associated with relatively high involvement of public bodies in auditing (developing, collecting evidence, judging evidence and judging auditors). Both axes are quantified averages of the factors in brackets, for a typical product or security audit for each industry, as of 2024. Here they are displayed as ranks along the axes, thus distances between points are not meaningful. Details in Appendix A. As of 2024, advanced AI auditing by public bodies (-appointed) is limited (Hobbhahn and Scheurer 2024). Criticality of advanced AI is unclear.

The case studies illustrate that auditing regimes strike different compromises between independence and efficiency. Variation in regime design relates to demand-side characteristics in each context, such as risk uncertainty, the costs of verifying the safety of the audited technology, and the sensitivity of information uncovered during the auditing process. These factors positively correlate with the public body assuming greater control over the auditing process, prioritizing independence, safety and public trust over efficiency.

For nuclear energy cybersecurity and aviation safety, we empirically find a "critical" risk profile, and a high involvement of public bodies. However, it is not always effective for the public body to be highly involved in auditing. Intuitively, the more auditing that is demanded (because, for example, the market is larger and there are more audited firms), the more challenging it becomes for the public body to conduct each and every audit. For example, the diversity and quantity of radio frequency devices constrain the ability of the public body to conduct auditing in every instance. Similarly, the regime for public firms' annual reports requires more efficient auditing. Potential harm by radio frequency devices or accounting is relatively low, audit information less sensitive and verification possible without extensive trials. Thus, private parties are responsible for most auditing steps.

The following figure illustrates a potential explanation for different auditing responsibilities. Industry conditions and audit conditions (demand-side factors) demand different auditor characteristics (supply-side factors), essentially determining whether independence or efficient capacity are more important, which dictates who audits (auditing responsibility).

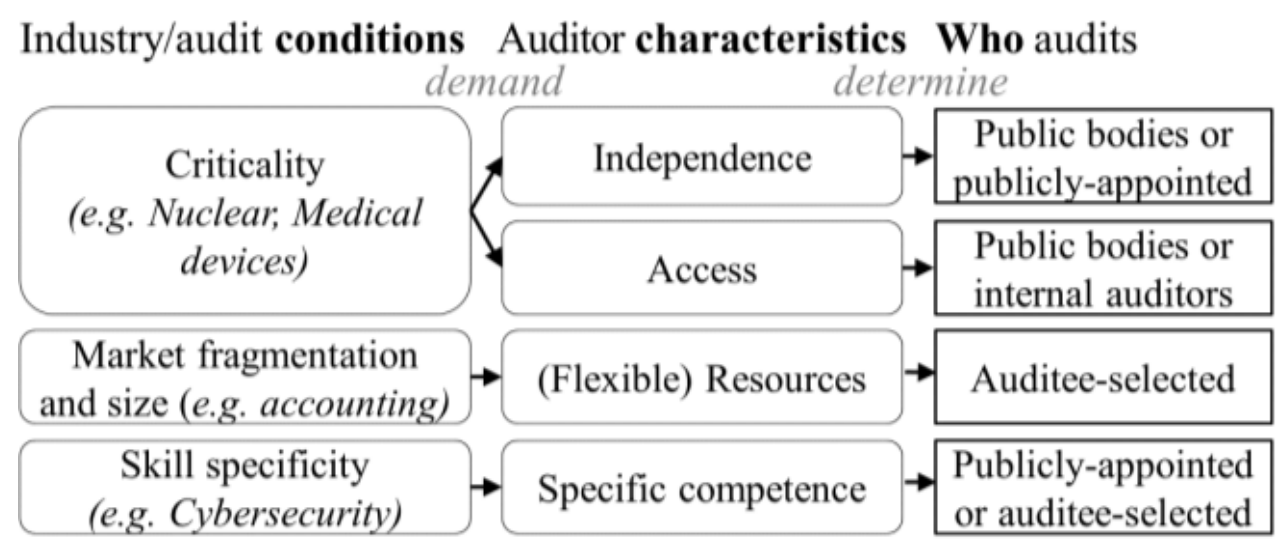

Figure 2. Connection between demand-side factors, supply-side factors and auditing responsibility. Note that auditor capacity can be influenced (see section 7). For each case, a combination of criticality, market concentration and skill specificity influences who audits, while criticality seems most prominent.

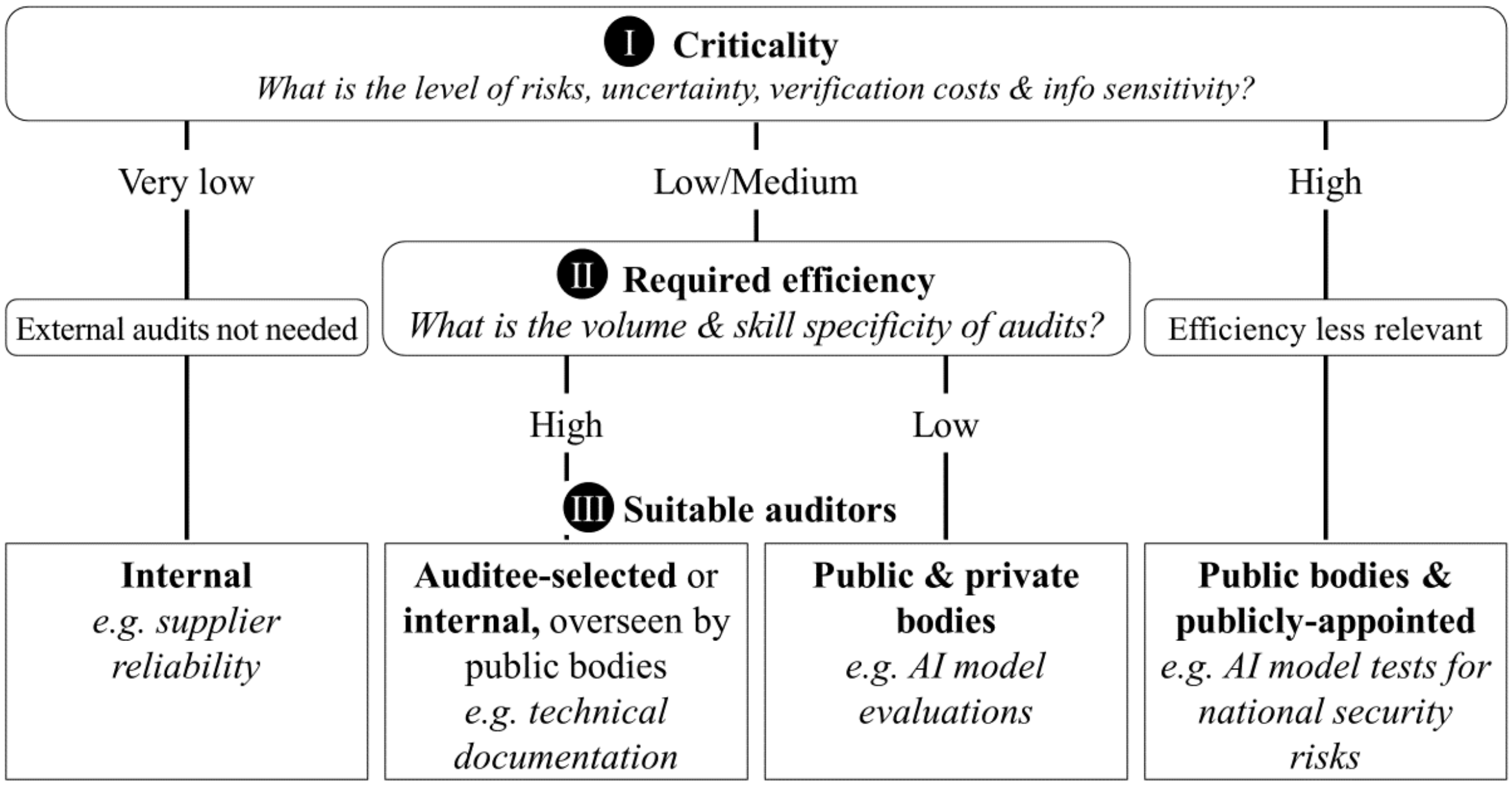

Figure 3: 3-step decision logic for running advanced AI auditing. Suitable auditors are indicative for collecting and judging evidence. The suitability is based on case study evidence on criticality and efficiency, and qualitatively explainable with auditor characteristics of independence and resources. Most likely, all auditor types might be involved in developing audits. AI Labs might support in all cases with collecting evidence. The volume of audits depends on future developments of market concentration and market size.

### 5.2 Three-Step Logic for Auditing Regime Design

Drawing on the quantification and analysis of cases above, we develop a three-step logic, intended to guide policymakers' auditing regime design choices (Figure 3).

- **Step 1 - Criticality**. Is the audit critical, necessitating an independent audit from a public body or publicly-appointed auditor? Criticality depends primarily on the risk level, risk uncertainty, verification costs and information sensitivity associated with the particular audit. It is only non-critical to involve auditee-selected auditors if the associated risks are well understood and the testing procedure is standardized.
- **Step 2 - Efficiency**. Who has or can efficiently build the required resources, competence and access? In this regard, we consider the volume of audits and the required skill specificity. If the volume of audits is high, and auditors do not require access to sensitive information, private parties may be tasked with auditing.
- **Step 3 - Suitable auditors**. Steps 1 and 2 determine which auditor characteristics are most demanded, and auditors with fitting characteristics are thus suitable.

This three-step logic is an idealized deduction from the case studies, reducing them to factors that previous literature and hybrid governance theory reasonably expects to influence audit effectiveness, as per our framework. However, the emergence of regimes is shaped by many other historical factors too, including political dynamics or concentration of skills in certain government departments (Ayres and Braithwaite 1992), as reviewed for each case in detail in Appendix A.1.

## 6 The Role of Public Bodies in an Advanced AI Auditing Regime

Public bodies can be involved in 6.1) different types of AI audits along different stages of the auditing lifecycle. The demand-side factors of each type determine 6.2) the optimal role of the public body in line with the three-step logic.

### 6.1 Advanced AI Audit Scope

There are many scopes or types of advanced AI audits. We focus on audits relevant to the development and provision of the AI model, and thus exclude product audits. We distinguish between those that focus on the governance practices of the firm that develops and provides advanced AI models, the security systems in place to prevent unauthorized access to the AI Lab's software and data, and the capability, alignment and sociotechnical impacts of an AI model (Moekander et al. 2023, EU AI Act).

**Governance audits** ensure the firm meets structural and procedural prescriptions (Moekander et al. 2023, Crawford 2022). Governance audits are predominantly qualitative and examine documentation concerning the auditee's:

- Risk management system: risk identification, assessment, thresholds and mitigations, with emergency protocols in case of major incidents (Barrett et al. 2023)
- Quality management system: roles and responsibilities, points of contacts, system architecture, data governance
- Data audits (Birhane et al. 2024)
- Ecosystem audits: environmental reporting, labor, supply chain (Birhane et al. 2024)

**Security audits** evaluate the robustness of systems that prevent unauthorized access to and use of an AI Lab's technologies and data. They encompass assessments of cybersecurity systems, physical security systems, and information security systems (Nevo et al. 2023, Huang et al. 2024, Alaghbari et al. 2022).

**Model audits** evaluate AI models to explain their behaviors, assess their capabilities, and test their capacity for harm in user interactions and sociotechnical impacts (Weidinger et al. 2023, Casper et al. 2024). Black-box evaluation techniques assess an AI model's performance from an external (e.g., user) perspective, limiting analysis to the model's inputs and outputs without accessing its internal workings (Casper et al. 2024). By contrast, white-box techniques involve analyzing the internal functioning of the model (Casper et al. 2024). Intermediate approaches are referred to as 'gray-box'.

The required comprehensiveness of an audit may scale with an AI model's capabilities. For example, highly capable models, such as those trained with substantial computational resources, may require more rigorous audits. Common tiers of model audits include but are not limited to (OpenAI 2023, Anthropic 2024):

1. *Single-shot or few-shot benchmarking*. Evaluating the model's performance on specific tasks such as answering a set of multiple choice questions. There are different suites of benchmarks including the 'Measuring Multitask Language Understanding' (MMLU) measures, which test model accuracy on '57 tasks ranging from mathematics to history to law' (Anthropic, 2024, Liang et al. 2023)
2. *Black-box adversarial testing*. Technique aimed at intentionally exploiting a model to produce not intended outputs, such as an offensive image or instructions for cyberattacks. This may leverage domain-specific expertise, such as knowledge of chemical, biological, radiological and nuclear ("CBRN") threats (Anthropic 2023).
3. *Gray- or white-box, or scaffolding-enhanced adversarial tests*. Elicitation of capabilities and propensities of model behavior with extensive tooling on-top of the model or analysis of the internals of the model (Anthropic 2023).
4. *Systemic impact evaluations*, including human interaction evaluations, systemic safety monitoring, sociotechnical user studies, uplift studies and yet-to-be-developed audits of specific societal areas (Weidinger et al. 2023, Stein and Dunlop 2024).

This typology is not exhaustive. Other kinds of audits relevant to advanced AI are emerging such as code inspections (Cohen at al. 2024) and audits of computational resources (Sastry et al. 2024).

## 6.2 The Public Body's Optimal Role in an Advanced AI Auditing Regime

Below we apply the logic developed in Section 5 to the advanced AI context (Detailed sources: Appendix A.2.).

**Demand-Side Analysis: Industry and Audit Factors**

<u>Industry conditions</u>

*Risk Uncertainty.* Advanced AI is a complex and evolving general-purpose technology with implications for users and external systems that are expanding and difficult to reliably estimate, i.e. highly uncertain (DSIT 2024). There is a record number of 12 related standardization requests under discussions in JTC 21.

*Potential for Externalities.* Advanced AI already proliferates rapidly, with hundreds of millions of users worldwide (Stein and Dunlop 2024). The generality leads to an indefinite number of potential downstream use cases. The degree of risk externalities is debated and uncertain. In some scenarios, advanced AI poses catastrophic risks, in others, rather low externalities.

*Public Salience.* Currently, public salience of advanced AI risk is high (as measured by Google News results, see Appendix A.2), which allows for the build-up of public oversight capacity.

*Market size and concentration.* As a technology with high returns to scale, advanced AI model providers are highly concentrated. The 2024 generative AI industry size is $25 billion in the US (Statista 2024b). The industry is growing, but the audit volume remains highly uncertain.

<u>Audit conditions</u>

*Verification Costs.* Verifying the risks, safety and compliance of advanced AI systems can be complex and potentially costly, depending on the audit scope (see Table 5, and Brundage et al. (2020), Casper et al. (2024)). Current methods for adversarial tests, systemic impact analysis and security audits are unstandardized and require significant expertise, time, and resources, making thorough verification challenging. Other methods, like benchmarking, are less time intensive and more standardized.

*Information Sensitivity.* Adversarial model audits that identify flaws and vulnerabilities in highly capable AI models are sensitive to the extent they reveal pathways to misusing advanced AI for harmful purposes, like cyberattacks or CBRN threats. There are also concerns that sensitive model test results could enter the training datasets of advanced AI. Due to the national security relevance of advanced AI, audits of security and AI models are sensitive. On the other hand, API-based black-box model evaluations need less sensitive information.

*Skill Specificity.* As foreshadowed in subsection 6.1, we suggest that particular AI model audits, as opposed to governance and security audits, require significant and specialized expertise. Domain-specific expertise is required to develop threat models and red-team advanced AI. Research engineers and computational social scientists are required to understand models and their impacts.

| **Audit scope** | **Demand-side factors**<br>Industry risk profile (→*Resources*) | Skill specificity (→*Competence*) | Verification costs & info sensitivity (→*Access*) |
|---|---|---|---|
| **Governance** | █ | ▒ E.g., Auditors in Compliance | █ E.g. Partly manual documentation |
| **Security** | █ | ▒ E.g., Security professionals | E.g. Inspections, partly manual |
| **Model...** | | | |
| ..Benchmarks | ▒ | █ E.g., ML engineers | E.g. Black-box, automated |
| ..Adversarial tests | ░ | ▒ E.g., Domain & ML experts | ▒ E.g. Grey-/White-box, manual |
| ...Systemic impact | ░ | ▒ E.g., Social scientists | ▒ E.g. Black-box / Usage, manual |

*Level of demand-side factors:* █ *High* ▒ *Medium* ░ *Low*

Table 5: Assumed status quo of demand-side factors by audit scope, for advanced AI. Industry risk profile includes risk uncertainty, potential for externalities and market concentration. Access from Casper et al. (2024); competence based on practitioner input (see Appendix A).

**Supply-Side Analysis: The Role of AI Safety Institutes and Other Public Bodies in Advanced AI Auditing**

An advanced AI auditing regime should be designed to incentivize an optimal balance between the auditor's independence, resources, competence and access to auditing evidence. Failing this, we expect auditing quality and its usefulness as a tool for monitoring regulatory compliance and the benefits and safety of AI systems to deteriorate. The demand-side analysis of the industry risk profile suggests that the unpredictable but potentially critical and far-reaching impacts of advanced AI justify the prioritization of independence and, consequently, public body involvement. However, this finding is complicated by intersecting efficiency challenges of using existing and building new competence, access and capacity in a nascent and unstandardized ecosystem for AI model audits. Such audits require niche expertise and innovation in auditing practices. Therefore, as illustrated by Table 5, we suggest that different aims and types of AI audits invite different auditing regime considerations.

**Implication 1: Public Body Involvement in Gray- and Black-Box Model Evaluations for Critical Risks**

If policymakers agree with the demand-side analysis, we suggest that the public body should be directly involved in certain kinds of advanced AI model audits that: (a) pertain to critical risks such as those affecting national security; (b) demand white- or gray-box access to AI models (such as certain kinds of adversarial tests and evaluations); and (c) involve access to sensitive information. This model loosely resembles auditing regimes in aviation and nuclear energy. In line with the three-step logic, suitable auditors for such high criticality tasks are public bodies & publicly-appointed externals. Given concentration in the advanced AI market and the prohibitive costs of training state-of-the-art models, the volume of audits might allow for such high involvement of less efficient public bodies. However, as discussed in subsections 7.1 and 7.2 below, the challenge for policymakers is to ensure the public body possesses adequate expertise and knowledge of the advanced AI system to conduct intensive, complex and potentially bespoke model evaluations (Casper et al. 2024, Anthropic 2023). This could manifest as an integrated team comprising government officials, publicly-appointed experts and senior representatives from the AI Lab itself.

**Implication 2: Public Oversight of an Auditing Market for Governance, Security and Select Model Audits**

Governance and security audits of AI Labs are more standardized, tap into auditing practices that are relatively mature in other industry contexts, and, apart from certain kinds of security audits, do not entail access to information that would harm the public if disclosed (Schuett 2023, Bos 2018). We suggest, therefore, that such audits could be provided by a market of private auditors supplemented by public body oversight. The public body's role should be to facilitate high quality auditing through policies that augment auditor independence and competence. These should include schemes to accredit auditor expertise and regulations that impose quality standards on auditors with consequences for failure.

These considerations may also extend to certain kinds of black-box model audits such as benchmark evaluations that assess AI model performance on standardized tasks. Such evaluations do not typically involve highly sensitive information and could benefit from the competitive dynamics of a private auditing market, generating innovation and expertise in AI auditing practices.

## 7 Public Body Capacity Estimates

As a consequence of the criticality of some advanced AI audits and the necessity for government involvement analyzed in the previous chapter, public bodies like AI Safety Institutes must build regulatory capacity, technical competence and ensure information access to both conduct certain kinds of audits and oversee others. In this section, we estimate the resources, competence and access requirements of the public body, with reference to case study evidence. Shortfalls in public bodies' capacity, competence and access limited effective AI auditing in the past (Lawrence et al. 2023, Groves et al. 2024, Politico 2024). Our figures are estimates only and assume the public body is operating in an advanced economy with a remit covering the current size of the advanced AI industry in the US.

When audits are critical: **More FTE at public bodies**

Criticality & market concentration (increasing upward)

| | # FTE | # FTE (scaled) | Technical FTE | Info access |
|---|---|---|---|---|
| Cyber (Nuclear) | 40 | 1250 – 2000 | 40-60% | On demand |
| Medical devices | 1200 | 100-200 | 70-90% | On demand |
| Aviation | 1800 | 600–900 | 70-90% | On demand |
| **Advanced AI** | | ? | ? | ? |
| Cyber (Power grid) | <30 | 250-400 | 70-90% | If suspicion |
| Finance (Securities) | 30 | <30 | 0-20% | Limited |
| Telco (Devices) | <30 | <30 | 20-40% | Limited |
| Accounting | <30 | <30 | 70-90% | If suspicion |

Figure 5: Public bodies' resources across case studies in the US, sorted by criticality and market concentration. FTEs (Full-time equivalents) are scaled to the current advanced AI industry size of $25 billion in the US (Statista 2024b). Share of technical FTE are roles framed as "specialists". "Supportive" roles are non-technical staff. Info access on demand for random inspections (See details in Appendices A.3, B.4 and B.5)

## 7.1 Resources

**Case Study Evidence**

The cases suggest that in auditing regimes where the public body is directly involved in auditing, the public body employs more staff relative to when the public body is an overseer of private auditors. As analyzed previously, the public body is more involved in auditing, when criticality and market concentration are high. Thus, higher criticality and market concentration demands more staff at public bodies, as shown in Figure 5.

**Implication 3: 100s of FTE for Advanced AI Auditing**

For effective advanced AI auditing, public bodies' auditing-related FTEs, share of technical staff and access, will need to be roughly on par with public bodies active in other industries with similar criticality and market concentration. If criticality and market concentration of advanced AI remains high and thus demands high public involvement in auditing, then the public body will need 100s of auditing FTEs in jurisdictions like the US.

Advanced AI models or security audits that entangle sensitive information require precautionary measures to ensure evaluation results or test sets are not leaked publicly or introduced into the AI model's training data. We assume that particular model or security audits will be resource-intensive with up to a dozen auditors being required to collect and elicit evidence in respect of a single threat (see Appendix A.3 and B.4 for detailed estimations for each audit method).

A surge in new AI Labs, models and risks will require the public body to increase its auditing capacity. To adapt to changes in demand, public bodies may need to develop organizational slack (Bourgeois 1981) or flexibility by, for example, maintaining and drawing on a pool of accredited AI auditing experts from academia or private sectors. Framework agreements could assist in accelerating their appointment.

## 7.2 Competence

**Case Study Evidence**

What kind of staff are needed? We find that in auditing regimes where the public body is directly involved in specialist auditing methods related to a complex product or technology rather than corporate governance, a higher proportion of the public body's staff are technical specialists. For example, auditing teams in the US Food and Drug Administration ("FDA") are composed of >70% technical specialists, which reflects that the FDA is directly involved in assessing complex products such as medical devices. In regimes where the public body oversees a private auditing market, the public body is able to develop more generalist competence to, for example, focus on assessing auditors and processes rather than the safety and benefits of a technology itself.

**Implication 4: Extensive, Diverse Technical Expertise at Public Bodies to Verify Claims of AI Labs**

Required staff skill profiles vary depending on the specific audit. Public bodies require a mixed technical and non-technical team dedicated to developing, conducting or judging audits. This team should involve a mix of computer engineers, compliance specialists, and domain-specific experts from fields like cybersecurity (See Table 5). The share of technical profiles will depend on the public body's degree of involvement in auditing. As discussed, it seems likely that the public body will be very involved, at least in the medium-term, given high levels of risk uncertainty and a lack of standardization.

## 7.3 Access and Learning

**Case Study Evidence**

In addition to capacity and competence, auditors require access to the information necessary for auditing. When risk uncertainty and verification costs are high, public bodies and appointed auditors need extensive access to information held by auditees and auditors (Costanza-Chock et al. 2023).

Not only does sufficient access to information underpin auditing quality, it may also facilitate the development of auditing competence and standardization (Schelker 2010). However, private players who learn the most through internal access may not always have the incentive to share their learnings - as seen in the case of oil companies research on climate change or financial auditors' withholding of information as part of the Enron scandal (Petrick & Scherer 2003). Therefore, public bodies should ensure their learning through mandating access to: (A) auditees' information; and (B) auditors' information.

For technical, profit-aligned developments of audits, firms share information, speed up standardization and, in turn, increase innovation - like for telecommunications and cybersecurity in bulk power systems (Blind 2013, Blind 2006). In healthcare, cyber for government contractors, aviation, cyber for nuclear and life sciences, public bodies learned through continued information access, enabling more standardized guidelines and, over time, auditing by private auditors instead of directly by public bodies.

**Implication 5: Structured Access to Auditee and Auditor Information**

Verifying claims and conformance with rules of AI Labs requires structured access to facilities, security systems, and the AI model. Lacking access is a noted challenge for AI auditors (Costanza-Chock et al. 2023, Casper et al. 2024). To effectively collect and judge evidence, e.g. by conducting in-depth evaluations and adversarial tests, gray- and white-box access might be required (see Figure 6). For API-based benchmarks or developing audits access to proxies and analogous samples (i.e., sufficiently similar but not identical datasets) may suffice. Systemic impact and human interaction evaluations might require access to anonymized usage or human trial data (Weidinger et al. 2023).

Given the current concentration of expertise and the need to swiftly develop (harmonized) standards in advanced AI, public bodies and trusted researchers need access to private sector expertise and information.

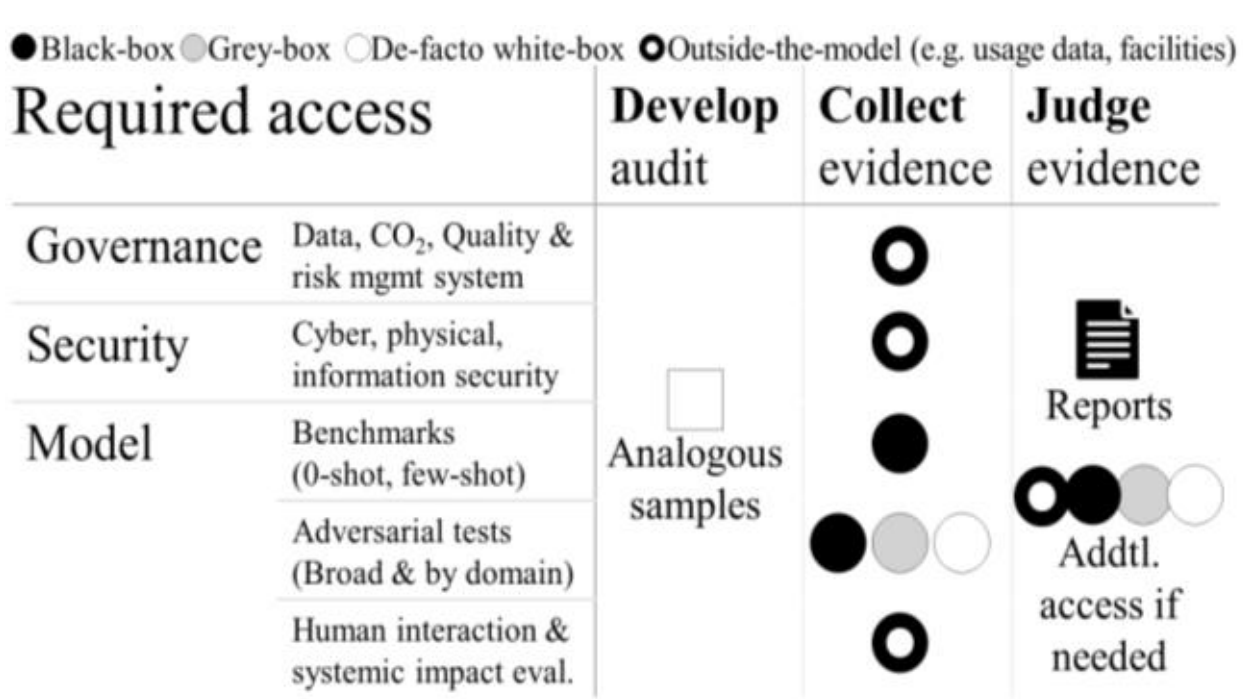

Figure 6: Access for auditing advanced AI. Terminology based on Casper et al. (2024)

When risks are more certain and audits standardized, auditee information can be restricted to cases of suspicion. AI audits that identify flaws and vulnerabilities in highly capable models may reveal pathways to misusing advanced AI for harmful purposes. Consequently, policymakers must mandate the optimal level of information access for AI auditors, instigating safeguards such as the requirement to obtain security clearances for gray- and white-box audits of highly capable AI models.

## 8 Conclusion

Drawing on our analysis of auditing regimes across high-risk industries, we derived five implications for designing advanced AI auditing regimes. Implications 1 and 2 revealed that when advanced AI risks, risk uncertainty, verification costs and information sensitivity are at levels comparable to the nuclear energy or aviation sectors, public bodies and publicly-appointed specialists need to audit AI Labs directly. Implications 3, 4, and 5 described the required resources, competence and access for AI Safety Institutes and public bodies to fulfill their auditing role. In case of high criticality, 100s of sociotechnical FTE and structured access to auditee and auditor information are needed.

Future research could explore a wider range of auditing regimes and country contexts, using deductive methodologies to test findings. Such research might, for example, consider:

- Quantitatively investigating causal links between auditing regime design choices and regime effectiveness.
- Qualitatively describing nuanced dynamics within and between AI auditor organizations (both public and private), exploring, for example, power dynamics, regulatory capture, and cultural differences.
- Understand the historical political and institutional reasons how different regimes and public bodies developed best practices and standardized audits.
- Define how auditing intersects with other AI governance mechanisms as part of a comprehensive regime.

## Acknowledgments

We are grateful for helpful discussions and feedback from Aaron Maniam, Alan Chan, Alexis Carlier, Clíodhna Ní Ghuidhir, Elliot Jones, Friederike Grosse-Holz, Gaurav Sett, Herbie Bradley, Lewis Ho, Lisa Soder, Lujain Ibrahim, Patrick Levermore, Peter Wills, Roxana Radu and participants of multiple AI Governance workshops in Oxford.

# Appendices

## A. Comparative Case Study Analysis

### A.1 Auditing Responsibilities

The summary figure below and the following figures contain

- a graphical description of auditing responsibilities in each case study
- historical context focused on regime emergence
- strengths and weaknesses of the respective regime.

The underlying definitions for auditing responsibilities mentioned in the figures are:

- **Develop** (“Rules” on case study figures): Who defines and changes rules and standards for auditees (and auditors)?
- **Collect** (“Info access”): Who collects evidence? Related to: Who has what degree of information access to collect evidence
- **Judge** (“Audit”): Who judges the collected evidence?

Further dimensions (In the case studies but not in the main text due to space constraints)

- **Auditing the Auditor** (“Audit”): Who judges auditing quality? Auditing the auditor refers to whether auditors are monitored by an external body.
- **Transparency**: Who can see which part of the audit results? Transparency refers to whether AI audit results are publicly disclosed.
- **Enforcement**: Who enforces consequences of non-compliance for auditees and auditors?

**Main sources per industry are the following: Accounting:** PCAOB (2024), Securities Exchange Commission (2003)

**Telco:** Federal Communications Commission (2024a, 2024b, 2024c, 2024d), Code of Federal Regulations Title 47 (2024), Code of Federal Regulations Title 47 (2024), National Institute of Standards (2024a), Maynard (2014), Hazlett and Pai (2018), Paglin (1989), European Commission (2018)

**Finance**: Securities and Exchange Commission (2020a), Financial Industry Regulatory Authority (2024), Financial Industry Regulatory Authority (2010), White (2018), S&P Global Ratings (2022), Securities and Exchange Commission (2020), Latham & Watkins (2022), Government Accountability Office (2021), Ryan (2012), Dennis (2008), Rivlin and Soroushian (2017), Lennon (2021), European Court of Auditors (2015)

**Aviation**: ISO/IEC 17021-1 (2015), NQA (2024), NQA (2024), Code of Federal Regulations Title 14 (2024), Federal Aviation Administration (2022), Federal Aviation Administration (2021a, 2021b, 2021c), Solomon (2010), Pasztor (2023), Neuman (2008), Davenport (2023), European Union Aviation Safety Agency (2024)

**Cyber-Power Grid:** ISO/IEC 27001 (2022), Federal Energy Regulatory Commission (2023c)

**Cyber-Government Contractors**: ISO/IEC 27001 (2022), Department of Defense (2024)

**Large Online Platforms**: Digital Services Act (2022)

**Life Sciences**: Food and Drug Administration (2024a, 2024b, 2024c, 2024d, 2024e), Stein and Dunlop (2023)

**Cyber-Nuclear**: ISO/IEC 27001 (2022), Nuclear Regulatory Commission (2024a).

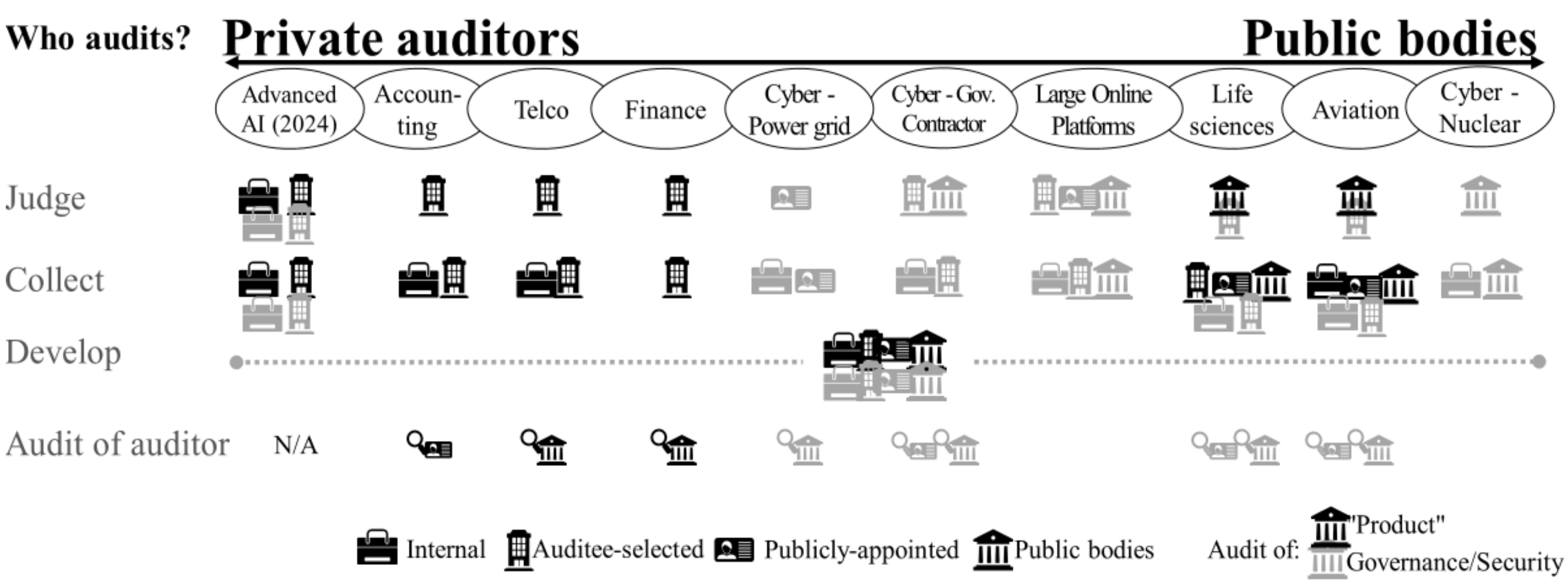

Figure A.1: Distribution of auditing responsibilities by auditor type for the US (EU for Online Platforms) as of 2024, based on case studies in the figures A.2-A.10.This excludes one-off audits without continuous access, as done by independent civil society organizations across industries and AI Safety Institutes (Politico 2024). Advanced AI regulation is changing fast, e.g., Similarly, AI Safety Institutes audits of advanced AI might develop into continuous audits beyond voluntary commitments

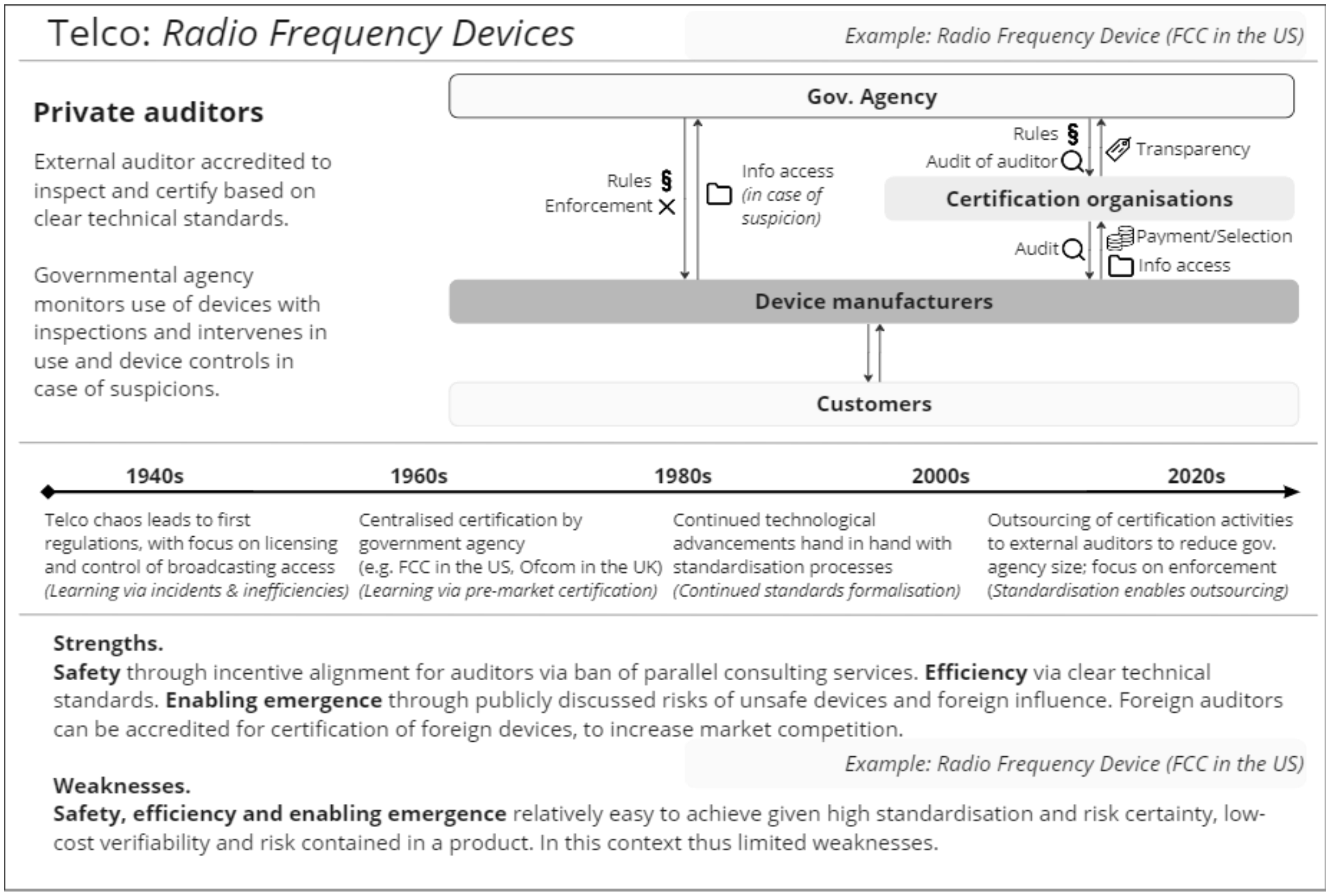

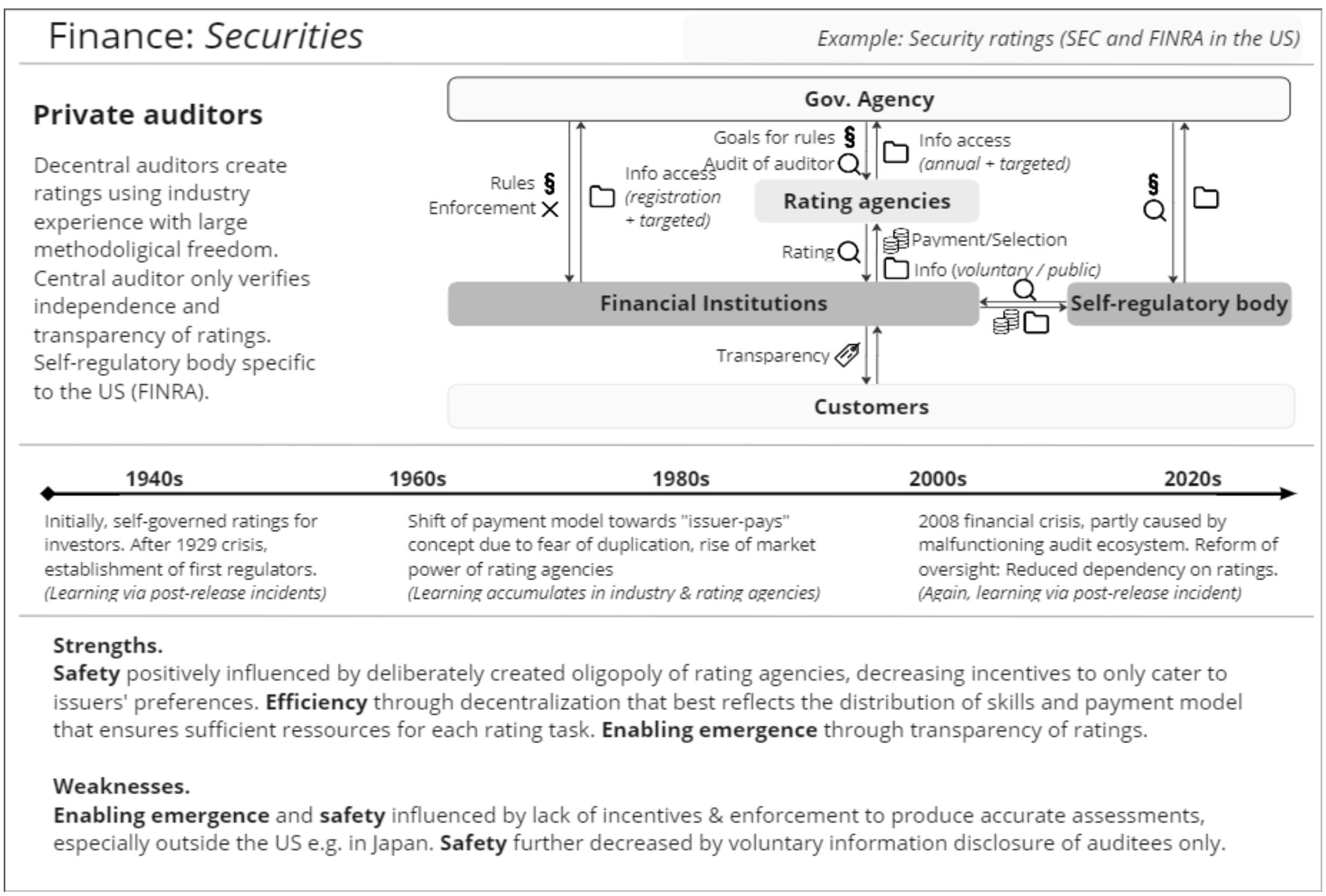

Figure A.2-3: Case study summary (Telco and Finance)

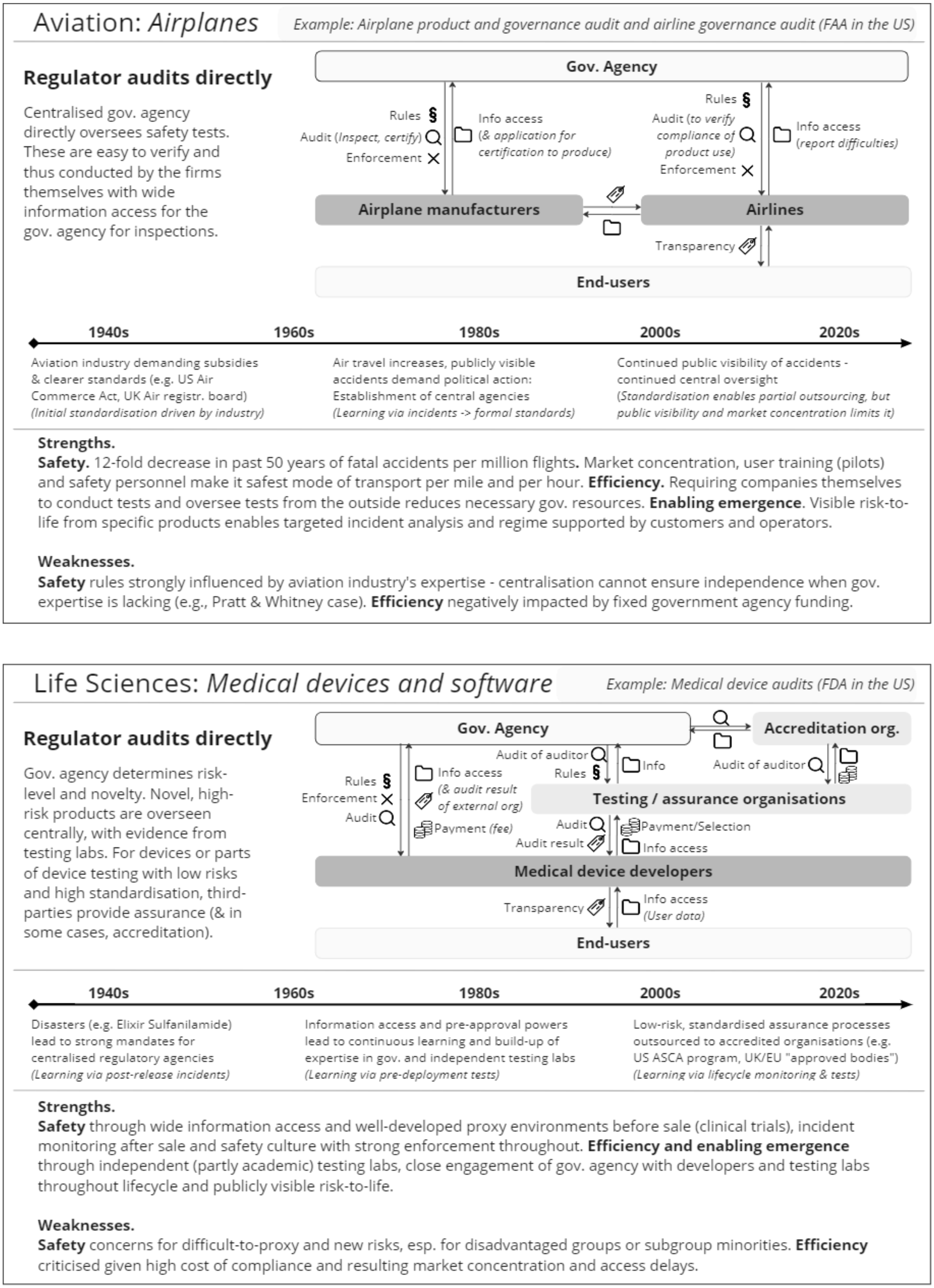

Figure A.4-5: Case study summary (Aviation and Life Sciences)

## Accounting: *Annual reports*

*Example: Annual reports of publicly listed companies (SEC and PCAOB in the US)*

### Private auditors

High concentration of skills, risks and power in auditors who are thus overseen tightly. Continuous discussions on the balance between independence and expertise of auditing bodies

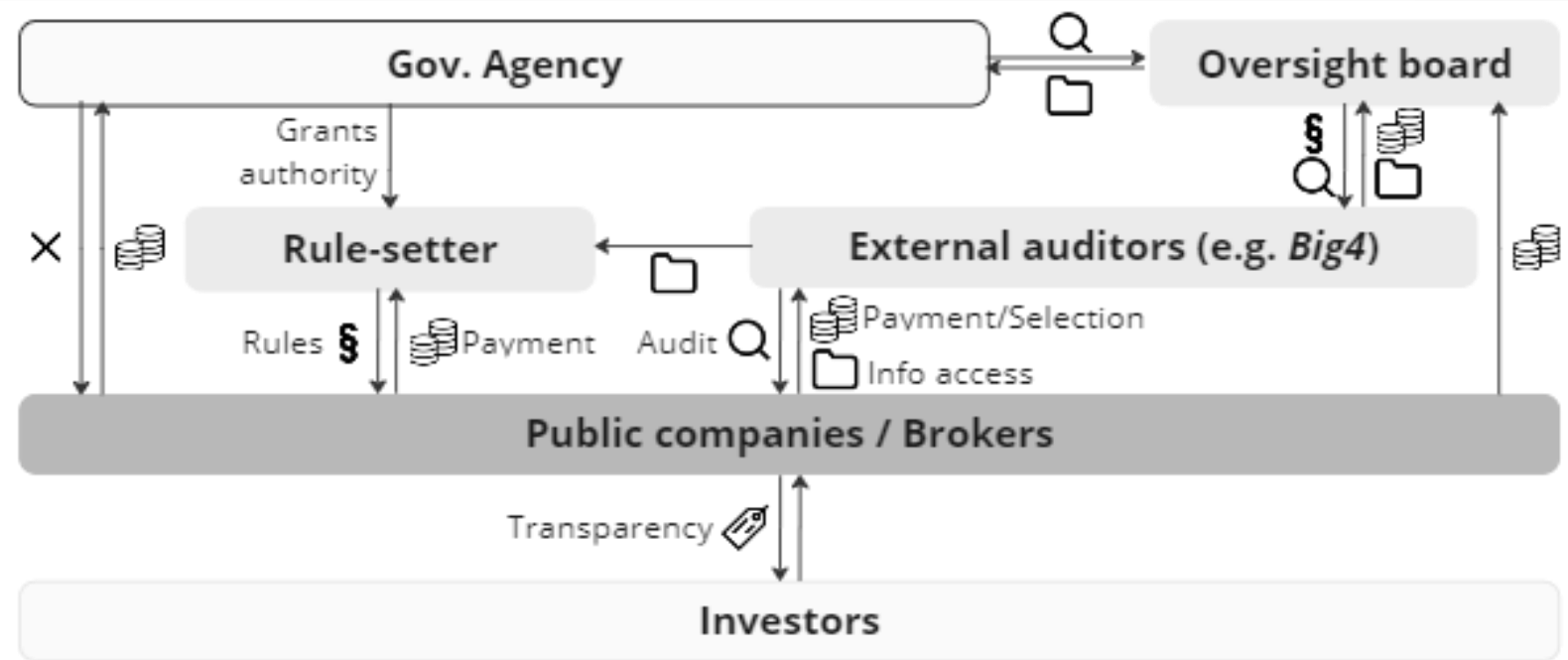

| 1940s | 1960s | 1980s | 2000s | 2020s |
|---|---|---|---|---|
| Investor-demanded oversight and accounting standards; establishment of gov. agencies *(Standards driven by customers)* | Self-regulation and growth of large accounting firms that accumulate expertise *(Learnings accumulates in auditors)* | Lacking gov. know-how limits gov. oversight: Continued self-regulation (e.g. FASB) *(Limited sharing of learnings)* | | In response to Enron scandal, stricter oversight on external auditors (e.g. PCAOB created) *(Again, learning via incidents)* |

**Strengths.**
**Efficiency** high due to concentration of expertise in auditors and within-firm internal auditing, and continuous standardisation for high verifiability. **Enabling emergence** through strong stakeholder support.

**Weaknesses.**
**Safety** continuous conflicts of interests, limited effect of auditor oversight beyond larger scandals due to limited public attention to niche issues, and concentrated expertise.

## Cyber: *Cybersecurity of nuclear energy plants*

*Example: Continuous Cyber tests (NRA in the US)*

### Regulator audits directly

Direct oversight due to national security relevance, information hazards and high, concentrated risks. Strong influence of industry associations (e.g. Nuclear Energy Institute) on rules, given its scientific focus. Information hazards limit public transparency.

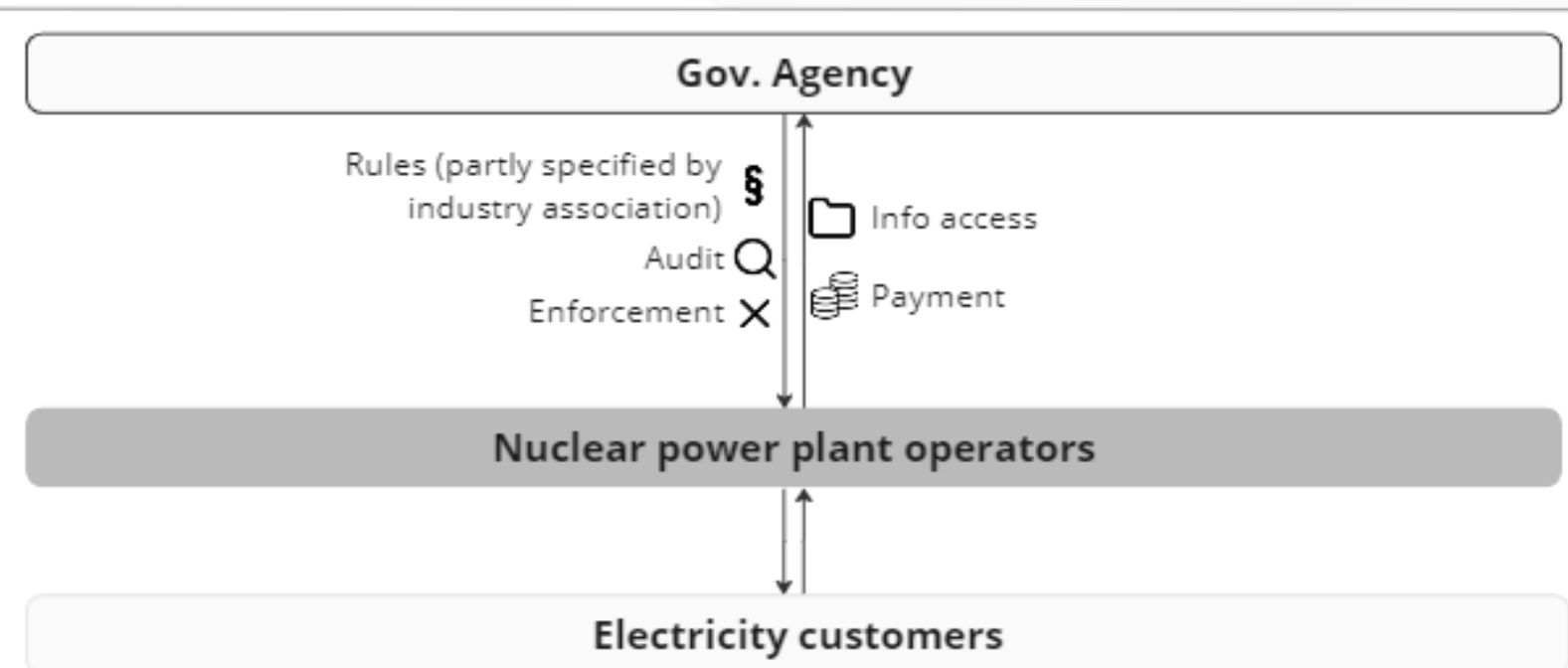

| 1940s | 1960s | 1980s | 2000s | 2020s |
|---|---|---|---|---|
| Military use of nuclear fission, high capital investments *(Learning via pre-deployment experiments)* | Initial governmental agencies focus on promotion of nuclear energy *(Standards driven by innovation needs)* | Safety-focused NRC formed, publicly visible accidents lead to tighter and centralised oversight (*Learning via incidents*) | | Beyond physical security & safety, Cybersecurity (& standards) added in existing regime (US: CFR 73.54) *(Learning via lifecycle monitoring & tests)* |

**Strengths.**
**Safety and efficiency** is high due to centralised regime, with clearly concentrated risks (2022: 411 Nuclear power plants worldwide). No publicly visible major cyber incident in nuclear facilities yet, given risk-based, strict separation of core and additional admin. systems.

**Weaknesses.**
**Enabling emergence.** Cybersecurity is an add-on to an existing, relatively static regime and security setup, which was not designed with cybersecurity in mind.

Figures A.6-7: Case study summary (Accounting and nuclear cybersecurity)

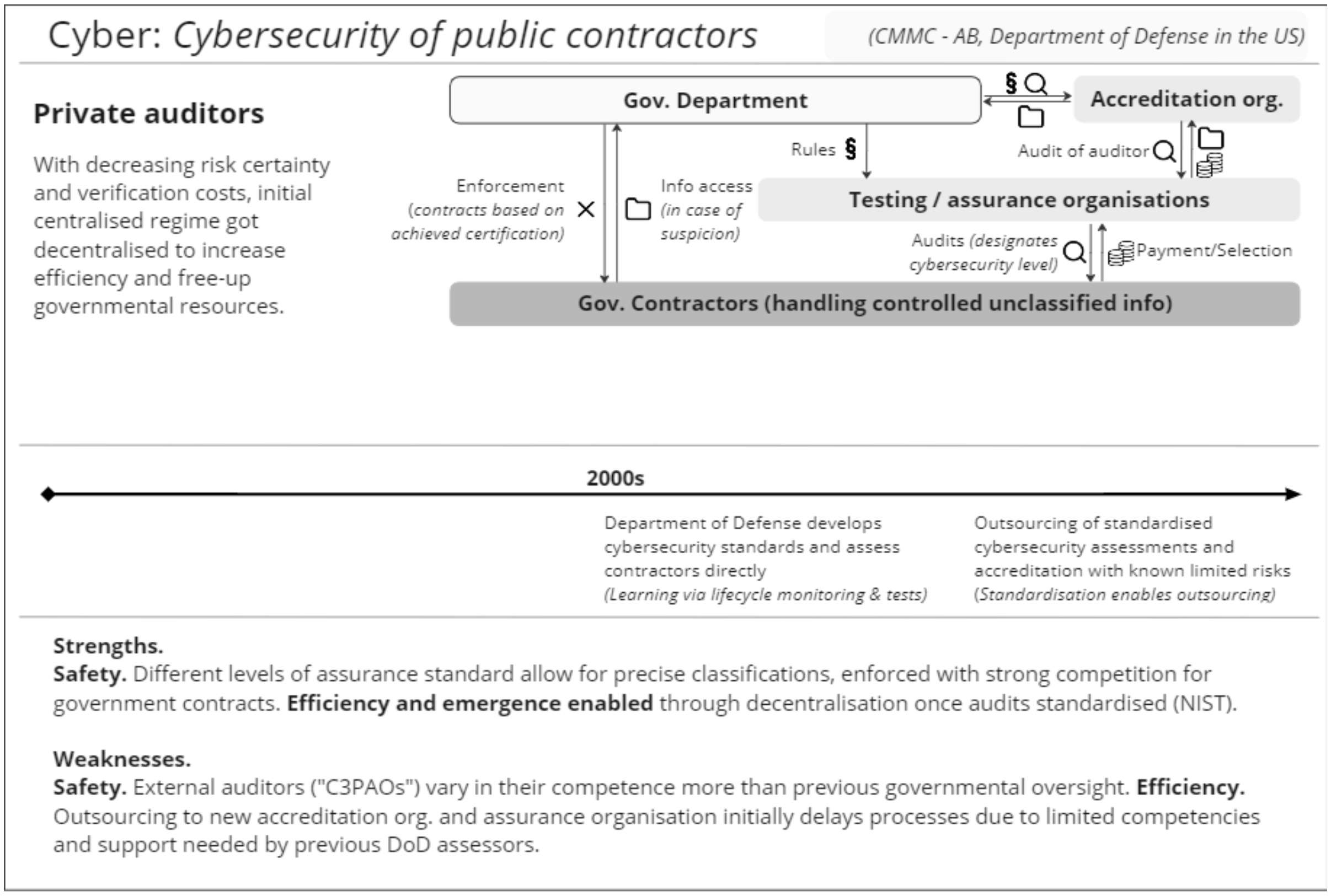

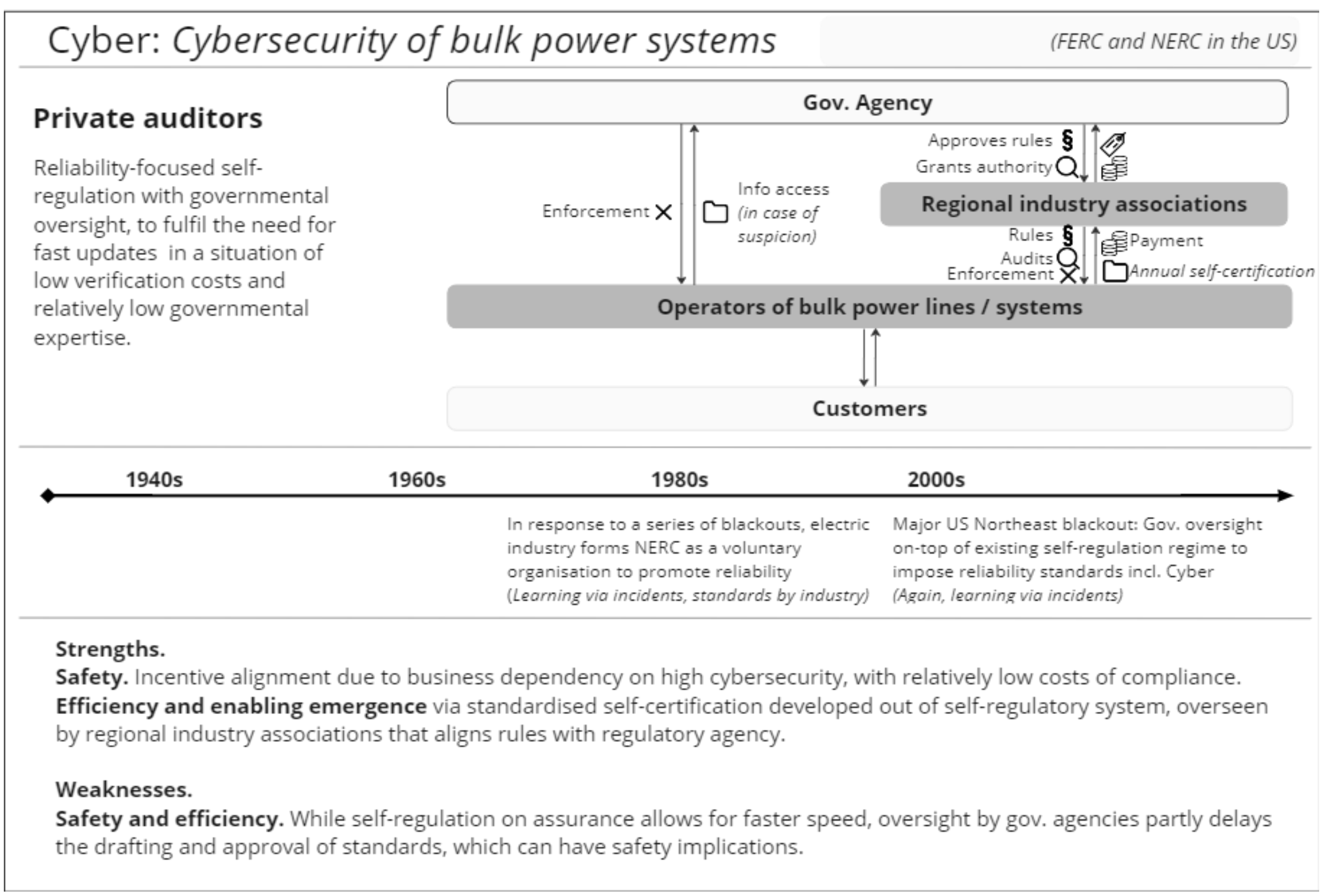

Figures A.8-9: Case study summary (Cybersecurity of contractors and of bulk power systems)

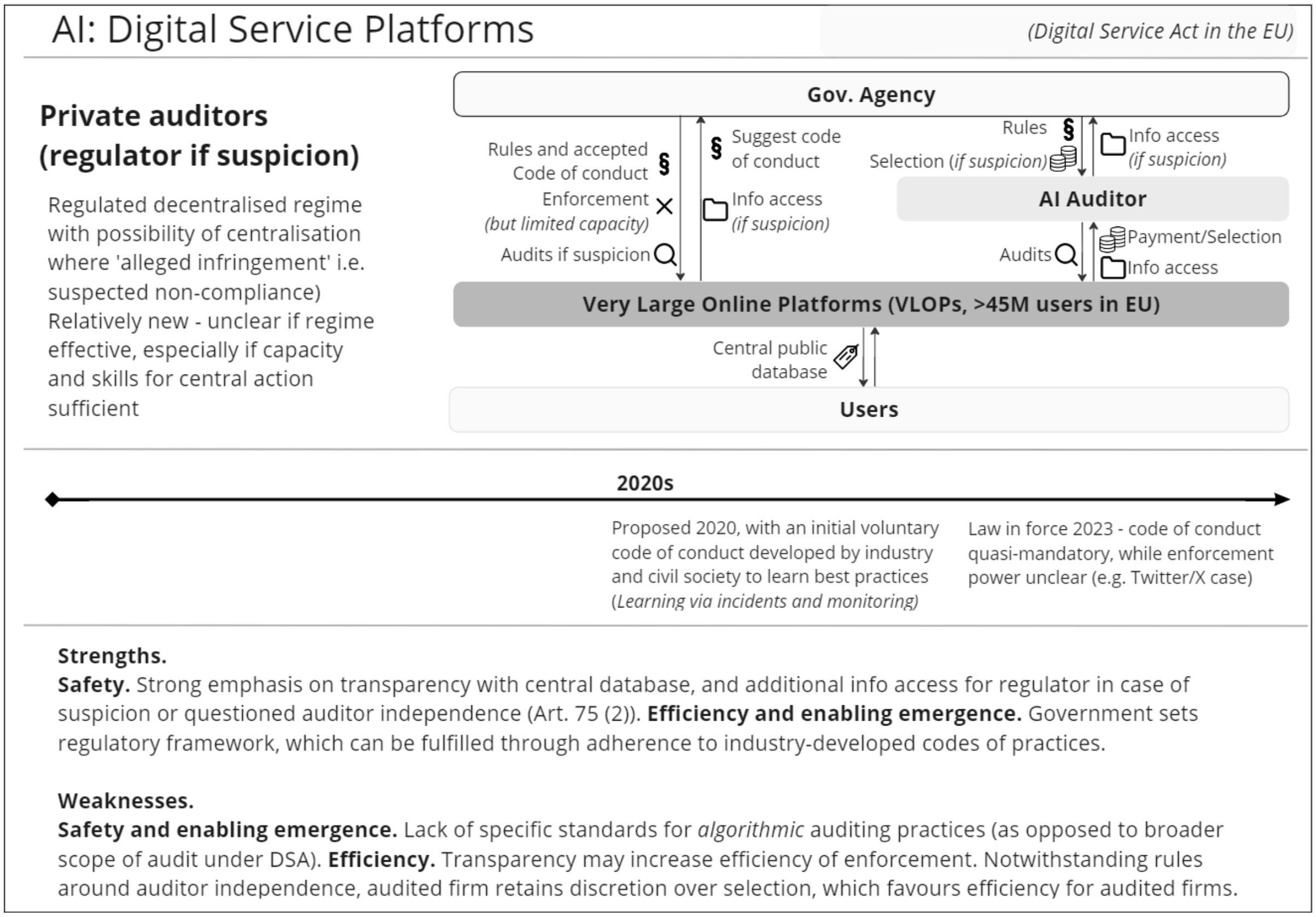

Figure A.10 Case study summary (Digital Service Platforms / Online Platforms)

### A.2 Demand-side factors

We determined the demand-side factors for each case study industry to uniformly characterize the respective environment of each case study industry.

The following tables contain (a) the average value across all demand-side factors for each case study industry (Figure A.11) as well as (b) the individual values for each demand-side factor and case study industry (Figures A.12 - 20). The underlying theoretical definition for each demand-side factor can be found in the background section of this appendix. The methods used to quantify and determine the demand-side factor per industry are described in the background section in Appendix B.1. While we estimate demand-side factors for advanced AI too, we span wide uncertainty given that is not an established industry like the others.

| Industry | # high | # medium | # low | AVERAGE |
|---|---|---|---|---|
| Accounting | 0 | 3 | 4 | **1.43** |
| Aviation | 3 | 4 | 0 | **2.43** |
| Life Sciences | 4 | 1 | 2 | **2.29** |
| Finance | 0 | 3 | 3 | **1.5** |
| Telco | 1 | 1 | 5 | **1.43** |
| Cybersecurity (Nuclear Energy) | 4 | 2 | 0 | **2.66** |
| Cybersecurity (Department of Defense Contractor) | 2 | 2 | 3 | **1.86** |
| Cybersecurity (Power Grids) | 2 | 2 | 3 | **1.86** |
| Advanced AI | 6 (max.) | 5 (min.) 1 (max.) | 2 (min.) | **1.7 (min.) - 2.86 (max.)** |
| Large Online Platforms | 2 | 2 | 2 | **2** |

Figure A.11: Average demand-side factors. Details from Appendix B.1: We used “1” for low, “2” for medium and “3” for high. These three chosen numbers have equal distances between each other by design, since “low”, “medium” and “high” were defined to have equal distance, too. We then calculated the average across these ordinal values for each industry.

| **Industry** | **Standardized value** | **Proxy Variable** | | | | | **Sources** |
|---|---|---|---|---|---|---|---|
| | | Experiments in use case environment | Onsite Inspection | Experiments in proxy environment | Simulation of use case environment | Outside logic verification (low) | |
| Accounting | **medium** | | X | | | X | see Case Study sources |
| Aviation | **high** | X | X | | | | see Case Study Sources |
| Life Sciences | **high** | X | X | X | | | MDC (2022) |
| Finance | **low** | | | | X | X | see Case Study Sources |
| Telco | **medium** | | X | X | | | see Case Study Sources |
| Cybersecurity (Nuclear Energy) | **medium** | | X (?) | | X | | Nuclear Regulatory Commission (2021c) |
| Cybersecurity (Department of Defense Contractor) | **low** | | | | X | | NIST (2023) |
| Cybersecurity (Power Grids) | **low** | | | | ? | | Nextlabs (2016) |
| Advanced AI | **medium (Could also be high)** | | X | X | X | | Brundage et al. (2020), Weidinger et al. (2023), Casper et al. (2024) |
| Large Online Platforms | **medium** | | X | | X | X | Digital Services Act (2022) |

Figure A.12: Verification Resources. Proxy Variable: Invasiveness of Test procedure

| **Industry** | **Standardized value** | **Proxy Variable 1** | | **Proxy Variable 2** | | | |
|---|---|---|---|---|---|---|---|
| | | Relevant ISO standards | Pages | sub TC | (i) under development | (ii) published | ratio (i)/(ii) |
| Accounting | **low** | ISO 5116-1/2/3:2021 | 106 | ISO/TC 68/SC 9 | 13 | 35 | 0.37 |
| Aviation | **medium** | 5 sub-TCs just on aerospace 345 standards | Average of 10 pages: 3,450 pages total | ISO/TC 20/SC: 1,4,8,9 ,10,17,18 | 28 | 338 | 0.08 |
| Life Sciences | **medium** | 2 sub-TCc all of which relevant to medical devices software (ISO/TC 194, ISO/TC 210)<br>67 standards | Average of 50 pages: 3,350 pages | ISO/TC 194, ISO/TC 210 | 29 | 67 | 0.43 |
| Finance | **medium** | Assessment based on changes in finance risk assessments before and after financial crisis | | | | | |
| Telco | **low** | 1 sub-TC (ISO/IEC JTC 1/SC 6), 405 standards | Average of 25 pages: 10,125 pages<br><br>Assume 50% relevant: Approx. 5,000 pages | ISO/IEC JTC 1/SC 6 | 21 | 405 | 0.05 |
| Cybersecurity (Nuclear Energy) | **medium** | Cybersecurity in general (JTC 1/SC 27)<br>40 standards<br><br>Nuclear safety: Approx. 5 dedicated standards within ISO/TC 85, e.g., ISO 7753:2023 | Average of 50 pages for cybersecurity<br>2,000 pages<br><br>Average of 30 pages für nuclear safety<br>150 pages<br>Aggregate: 2,150 pages | ISO/IEC JTC 1/SC 27<br>ISO/TC 85 | 117 | 503 | 0.23 |
| Cybersecurity (Department of Defense Contractor) | **low** | Cyber security more limited in scope than nuclear, probably 50% of standards relevant | 1,000 pages | ISO/IEC JTC 1/SC 27 | 72 | 240 | 0.30 |
| Cybersecurity (Power Grids) | **low** | Cyber security likely full scope. Possibly 29.240 or | 2,000 pages | ISO/IEC JTC 1/SC 27 | 72 | 240 | 0.30 |
| Advanced AI | **high (could also be medium)** | Based on 12 distinct standardization request under discussions in JTC 21 | | | | | |
| Large Online Platforms | **medium** | Based on extent of standardization requests during DSA law-making | | | | | |

Figure A.13: Risk Uncertainty. Proxy Variable 1: Total length of ISO standards, Proxy Variable 2: Share of standards under dev from total. Source: ISO standards mentioned above

| **Industry** | **Standardized value** | **Proxy Variable** | | | |
|---|---|---|---|---|---|
| | | Auditor industry (NAICS Code) | Score | Auditee industry (NAICS Code) | Score |
| Accounting | **low** | 541211: Accounting, Tax Preparation, Bookkeeping, and Payroll Services. | 138 | 541211: Accounting, Tax Preparation, Bookkeeping, and Payroll Services. | 500 |
| Aviation | **high** | 481: Air Transportation | 1030 | | |
| Life Sciences | **low** | 33911: Medical Equipment and Supplies Manufacturing | 191 | | |
| Finance | **medium** | Investment Banking and Securities Dealing: 523110 | 615 | 56145: Credit Bureaus (includes Credit Agencies) | 1192 |
| Telco | **low** | 334220: Radio and Television Broadcasting and Wireless Communications Equipment Manufacturing | 557 | | |
| Cybersecurity (Nuclear Energy) | **high** | 221113: Nuclear Electric Power Generation | 1500 | 541512: Computer Systems and Design Services (Cyber) | 176 |
| Cybersecurity (Department of Defense Contractor) | **high** | | | 541512: Computer Systems and Design Services (Cyber) | 176 |
| Cybersecurity (Power Grids) | **medium** | 2211: Electric Power Generation, Transmission, and Distribution | 211 | 541512: Computer Systems and Design Services (Cyber) | 176 |
| Advanced AI | **high (but depends on industry scenario, could also be medium)** | Assume 6 firms (like for DSA), 2 firms with 25% market share each, 4 firms with 12.5 share each. | 1875 (i.e. high) | | |
| Large Online Platforms | **high** | Assume 6 firms (see EU classification), 2 firms with 25% market share each, 4 firms with 12.5 share each. Inaccurate given submarkets but standardised classification likely ture. | 1875 (i.e. high) | | |

Figure A.14: Market concentration. Proxy: Herfindahl-Hirschman Index US for 50 largest firms (US Census Bureau 2017)

| Industry | Standardized value | Proxy Variable | | |
|---|---|---|---|---|
| | | job | Talent (k) | Glassdoor (k) |
| Accounting | **low** | Financial Reporting Accountant / Financial Accountant | 85 | 95 |
| Aviation | **medium** | Aircraft Engineer | 128 | 113 |
| Life Sciences | **high** | Medical Software Engineer | 130 | 171 |
| Finance | **low** | Credit Rating Analyst | | 105 |
| Telco | **low** | Telecommunication Engineer | 108 | |
| Cybersecurity (Nuclear Energy) | **medium** | Nuclear Cyber Security Analyst | | 122 |
| Cybersecurity (Department of Defense Contractor) | **medium** | Cybersecurity | 120 | 123 |
| Cybersecurity (Power Grids) | **medium** | Cyber Security Analyst | 120 | 123 |
| Advanced AI | **high (but depends on specific audit, could also be medium on avg.)** | Deep Learning Software Engineer | 150 | 182 |
| Large Online Platforms | **low** | Product Analyst | 97 | 82 |

Figure A.15: Skill Specificity. Proxy Variable: Average Salary. Sources: Talent (2024), Glassdoor (2024) for the US/EU.

| **Industry** | **Standardized value** | **Proxy Variable** | |
|---|---|---|---|
| | | industry term (+ risk) | search results, M |
| Accounting | **medium** | financial accounting | 14.5 |
| Aviation | **medium** | plane | 11.8 |
| Life Sciences | **high** | medical software | 20.7 |
| Finance | **medium** | securities rating | 14.1 |
| Telco | **low** | radio frequency device | 0.04 |
| Cybersecurity (Nuclear Energy) | **high** | cyber security nuclear plant | 43 |
| Cybersecurity (Department of Defense Contractor) | **low** | Cybersecurity contractors Department of Defense | 0.1 |
| Cybersecurity (Power Grids) | **low** | cybersecurity power grid | 0.05 |
| Advanced AI | **medium** | AI foundation model | 6.5 |
| Large Online Platforms | **high** | online platforms | 49 |

Figure A.16: Public Salience of industry risks. Proxy Variable: Number of Google News search results for <industry> risk, for 2019-2024

| **Industry** | **Standardized value** | **Proxy Variable** | | |
|---|---|---|---|---|
| | | most relevant risk event from report | impact (minor - catastrophic) | probabi-lity (%) |
| Accounting | **low** | none listed | minor | |
| Aviation | **medium** | aviation collision | significant | <0.2 |
| Life Sciences | **high** | Accident involving high-consequence dangerous goods | limited | 1-5% |
| Finance | **low** | Technological failure at a systemically important retail bank | moderate | 1-5% |
| Telco | **low** | none listed | minor | |
| Cybersecurity (Nuclear Energy) | **high** | Civil nuclear accident | catastrophic | <0.2 |
| Cybersecurity (Department of Defense Contractor) | **medium** | insolvency of supplier of critical services to the public sector | moderate | 5 to 25% |
| Cybersecurity (Power Grids) | **high** | Failure of the National Electricity Transmission System (NETS) | catastrophic | 1-5% |
| Advanced AI | **Depends on industry scenario: Could be Low or High** | none listed yet, but planned in upcoming versions | - | - |
| Large Online Platforms | **low** | Public disorder | limited | 1-5% |

Figure A.17: Scale of risk externality. Proxy Variable: National Risk Register (2023), and connected U.S. production account (U.S. Department of Labor 2012).

| **Industry** | **Standardized value** | **Proxy variable** | | | | **Sources** |
|---|---|---|---|---|---|---|
| | | **Public (= all)** | **Internal use only (= all within firm)** | **confidential (= legitimate interest within firm only)** | **restricted (= legitimate interest within firm only + further security tests)** | |
| Accounting | **medium** | | | x (to comply with customer information safeguarding requirements, see below) | | Code of Federal Regulations Title 16 (2024) |
| Aviation | **high** | | | | x (to ensure compliance with export control regulations as avionics regulated in EAR, see below) | Princeton University (2024)<br><br>ReedSmith (2024) |
| Life Sciences | **low** | x (minimal regulatory restrictions to share information) | | | | Food and Drug Administration (2024c) |
| Finance | **medium** | | | x (to comply with customer information safeguarding requirements, see below) | | Code of Federal Regulations Title 16 (2024) |
| Telco | **high** | | | | x (to ensure compliance with export control regulations as avionics regulated in EAR, see below) | Princeton University (2024)<br><br>ReedSmith (2024) |
| Cybersecurity (Nuclear Energy) | **high** | | | | x (to comply with SGI protection requirements and/or national security information protection) | The White House (2009)<br><br>NIST (2024b) |
| Cybersecurity (Department of Defense Contractor) | **high** | | | | x (to comply with security information protection, since infrastructure capability and vulnerability information is involved, could be all types of contractors therefore top secret level plausible) | Nuclear Regulatory Commission (2021a) |
| Cybersecurity (Power Grids) | **high** | | | | x (to comply with CEII requirements and/or national security information protection) | Federal Energy Regulatory Commission (2023b)<br><br>North American Electric Reliability Corporation (2023) |
| Advanced AI | **low (but depends on specific audit), could also be high on avg.** | x (limited regulatory restrictions to share information as of now, but likely changing in the future) | | | | The White House (2023) |
| Large Online Platforms | **medium** | | | x | | EU (2022) |

Figure A.18: Info sensitivity. Proxy variable: Strictest classification requirements (as mentioned in government documents that require compliance) of core information about product to be audited)

### A.3 Resource Analysis

We determined the number of employees (“resources”), the share of technical employees (“competence”) and the access to audit information (“access”) for the regulatory entity in each case study industry. These resource setups were then used to estimate the resource needs for the AI industry “top-down”. To verify this estimate for the AI industry, a “bottom-up” estimation was conducted.

Monitor Employees as a Percentage of
Combined Monitor and Lawyer Workforce[68]

| Light Monitors <15% | 15-49% | 50-85% | Heavy Monitors >85% |
|---|---|---|---|
| FTC 3% | FCC 34% | FERC 62% | FDA 98% |
| EEOC 0% | | EPA 60% | NCUA 97% |
| NLRB 0% | | CFPB 54% | FSIS 95% |
| | | SEC 53% | Fed. Res. 95% |
| | | | OSHA 93% |
| | | | NRC 93% |
| | | | FAA 93% |
| | | | FMCSA 93% |
| | | | OCC 93% |
| | | | MSHA 91% |
| | | | FDIC 86% |

Fig. A.19. In industries where the regulator audits, auditing or monitoring is a high share of workforce. Reproduced from Van Loo (2019).

#### Bottom-up estimation

All estimates represent 80% confidence intervals for both the current and ideal resources to run/judge an audit for one foundation model (or a significant update of one foundation model pre-deployment (e.g. Gemini, GPT3.5->GPT-4 or Claude 2->Claude 3))

Info access tiers are based on: Casper et al. (2024)

0 No access
1 Black Box
2 Grey Box
3 White Box
OtB Outside-the-Box

| | Goal | Task | Description | STATUS QUO Time & FTE (to run/judge) | In 1-3 years (If high criticality) FTE (to run/judge) | Access |
|---|---|---|---|---|---|---|
| Governance Audit | Collect evidence | Quality Management System (QMS) | Document roles & responsibilities<br>Document internal procedures (e.g. for review and sign-off )<br>Document procedure to handle user complaints<br>Document user manual for end users | 1 FTE for 2-8 weeks | 1-3 FTE for 2-8 weeks | 0 |
| Governance Audit | Collect | Risk Management System (RMS) | Document risks in likelihood-impact matrix<br>Document risk mitigation measures<br>Document emergency procedures | 1 FTE for 1-4 weeks | 1-2 FTE for 1-4 weeks | 0 |
| Governance Audit | Judge evidence | Evaluation & Recommendations | Evaluate all evidence and issue recommendations | <1 FTE for 1-2 weeks | 1 FTE for 1-2 weeks | 0 |
| Governance Audit | Overhead | Project management | Organise meetings, conduct interviews | 0.5-1 FTE for 2-4 weeks | 0.5-1 FTE for 2-4 weeks | N/A |
| Governance Audit | Overhead | Incorporate feedback | Update documentation based on audit findings | <1 FTE for 1-2 weeks | 1 FTE for 1-2 weeks | 0 |
| Data Audit | Collect | Technical Documentation | Describe data (sourcing strategy, quality assurance)<br>Evaluation of copyright infringements | <1 FTE for 1-4 months | 1 FTE for 1-4 months | *OtB* |
| Data Audit | Collect | Descriptive Analysis | Assess descriptive dimensions of data (e.g. representativeness, biases, privacy) | <1 FTE for 1-4 months | 1 FTE for 1-4 months | *OtB* |
| Data Audit | Judge | Evaluation & Recommendations | Evaluate all evidence and issue recommendations | <1 FTE for 1-2 weeks | 1 FTE for 1-2 weeks | 0 |
| Data Audit | Overhead | Project management | Organise meetings, conduct interviews | 0.5-1 FTE for 1 month | 0.5-1 FTE for 1 month | N/A |
| Model Eval | Collect | Technical Documentation | Document training and evaluation strategy, including human annotation strategy.<br>Document model specifications & design choices | 1-2 FTE for 1 month | 1-2 FTE for 1 month | *OtB* |
| Model Eval | Collect | Benchmarking (0-shot or automated) | Evaluate against benchmarks (SuperGLUE, BIG-Bench, select part of HELM)<br>Evaluate truthfulness (truthfulQA)<br>Evaluate fairness (some parts of DecodingTrust) | <3 FTE for 1 week | 1-3 FTE for 1 week | 1 |
| Model Eval | Collect | Benchmarking (non-automated few-shot) | Induce unwanted model behaviour through few-shot prompting (e.g. parts of HELM) | <3 FTE for 1 week | 2-5 FTE for 1 week | 1 |
| Model Eval | Collect | Adversarial testing (Expert red-teaming) | CBRN experts interact with model over long period of time to look for specific capabilities in a model for specific risks, Exploratory / Qualitative | <10 FTE for 1 month | 10-100 FTE for 1 month | (1 or 2)? |
| Model Eval | Collect | Capability or propensity elicitation with specific finetuning or scaffolding environments | Models Autonomous replication evaluation (METR 2024), Apollo's deception evals, CBRN scaffolding based evals | <10 FTE for 1 month | 10-100 FTE for 1 month | 3 |
| Model Eval | Collect | Human-interaction evaluations and monitoring | Behavioral experiments, monitoring of human use (As defined in Weidinger et al. 2023) | <5 FTE for 1 month | 10-100 FTE for 1 month | 1 , 2 or 3 |
| Model Eval | Collect | Systemic impact evaluations | Impact Assessments, Pilot studies, Simulations (As defined in Weidinger et al. 2023) | 1-5 FTE for 1 month | 10-100 FTE for 1 month | 1 or OtB |
| Model Eval | Judge | Evalation & Recommendations | Evaluate all evidence on model evals and issue recommendations | <2 FTE for 1 month | 3-20 FTE for 1 month | 0 |
| Model Eval | Overhead | Project management | Organise meetings, conduct interviews, track various tests | <2 FTE for 1 month | 2-5 FTE for 1 month | N/A |
| Cybersecurity Audit | Collect | Tests and Procedures | Penetration Testing<br>Vulnerability Assessment<br>Incident Response Testing | <5 FTE for 1 month | 2-5 FTE for 1 month | 0 |
| Cybersecurity Audit | Judge | Evalation & Recommendations | Evaluate the results and documentations of all tests; issue recommendations | <3 FTE for 1 month | 1-3 FTE for 1 month | 0 |
| Cybersecurity Audit | Overhead | Project management | Organise meetings, conduct interviews | 0.5-1 FTE for 1 month | 0.5-1 FTE for 1 month | N/A |

Figure A.20: Bottom-up estimation of resources and access required for sub-parts of Advanced AI audits. The methodology for the estimates is explained in Appendix B.4. Note: The field and best-practices are changing fast, estimates as of H1 2024. For more up-to-date estimates, contact the authors. OtB = Outside the box (i.e. not model-related)

**Top-down estimation: Resources & Access:**

| Industry | Source | Page (and relevant content on that page) | Relevant programmes and/or units (and FTE where available) | Tasks within programme or unit (and FTE where available) | Total | Potential assumptions (number always rounded up to avoid underestimation) | Total per billion dollars in revenue | Source for revenue (with comments about extraction of numbers) |
|---|---|---|---|---|---|---|---|---|
| Accounting | U.S. Securities and Exchange Commission (2023) | 43 for SEC numbers & tasks | “SEC Office of the Chief Accountant” **OoCA** (2) | **OoCA:**<br>> “U.S. Auditing Regulator (PCAOB) Board Appointments” (1)<br>> “U.S. Auditing Regulator Budget and Accounting Support Fee Approval” (1) | **2** | - | **ca. 0** | Extracted prediction for 2023 = 145 billion (Statista 2023a) |
| Aviation | Federal Aviation Administration (2021a) | 13 - 17 for tasks & numbers | “Flight Standards - **FS**” (5140)<br>“Aircraft certification - **AC**” (1354) | **FS**:<br>“Certification, inspection, surveillance, investigation, and enforcement | **1822** | **FS**:<br>> Resources equally distributed among three areas of responsibility mentioned -> ⅓ * | **28.92** | Consider that US contributing 46% of 321 billion globally, among which 43.5% were generated by the commercial aircraft |

| Industry | Source | Page (and relevant content on that page) | Relevant programmes and/or units (and FTE where available) | Tasks within programme or unit (and FTE where available) | Total | Potential assumptions (number always rounded up to avoid underestimation) | Total per billion dollars in revenue | Source for revenue (with comments about extraction of numbers) |
|---|---|---|---|---|---|---|---|---|
| | | | | activities" (1370)<br>**AC**:<br>> "Assuring design, production, and airworthiness certification programs comply with prescribed safety" (ca. 226) standards"<br>> "Providing oversight of production approval holders, individual designees, and delegated organisations" (ca. 226) | | 5140 = 1713 for "Certification, inspection, surveillance, investigation, and enforcement activities";<br>> Within "Certification, inspection, surveillance, investigation, and enforcement activities" equally distributed among responsibilities -> without enforcement: ⅘ * 1713 = 1370<br>**AC:** FTEs in division evenly distributed across the six tasks -> 1/6 * 1354 = 226 | | segment -> 321 * 0.45 * 0.435 = 63 billion (Precedence Research 2023) |
| Life Sciences (Medical devices) | Food and Drug Administration (2018)<br><br>Food and Drug Administration (2024b) | 1 for high-level number | "Center for Devices and Radiological Health - CDRH" (1887) | **CDHR**: Office of Product Evaluation and Quality (1207) | **1207** | **CDHR:** FTE proportional to FTE on leadership level per office as indicated in office overview (see source 2) -> 0.64 * 1887 = 1207 | **6.54** | Extract prediction for 2024 = 184.61 billion (Fortune Business Insights 2024) |
| Finance (Securities) | U.S. Securities and Exchange Commission (2023) | 51 | "SEC - Office of Credit Ratings - **OoCR**" (47) | **OoCR**:<br>> Examinations (5)<br>> NRSRO Registrations — Filed Applications, Amendments, Withdrawals, and Cancellations (25) | **30** | **OoCR**: Scaled sum of workload data must equal number of FTEs -> 10/95 * 47 FTE for examinations, 50/95 * 47 FTE for NRSRO registrations | **0.19** | Extract prediction for 2024 = 160.8 billion (Statista 2024a) |

| Industry | Source | Page (and relevant content on that page) | Relevant programmes and/or units (and FTE where available) | Tasks within programme or unit (and FTE where available) | Total | Potential assumptions (number always rounded up to avoid underestimation) | Total per billion dollars in revenue | Source for revenue (with comments about extraction of numbers) |
|---|---|---|---|---|---|---|---|---|
| | | | | | | | | |
| Telco (Radio Frequency devices) | Federal Communications Commission (2023)<br><br>Federal Communications Commission (2015)<br><br>Federal Communications Commission (2024a)<br><br>National Institute of Standards and Technology (2024c)<br><br>National Institute of Standards and Technology (2023) | Landing page, OoEaT Org Chart<br><br>List of responsibilities of laboratory division<br><br>High-level number of FTE at OoEaT<br><br>Landing page, number of employees at NIST NVLAP<br><br>Landing page, list of programme administered by NVLAP | "FCC-Office of Engineering and Technology - **OoEaT**" (79)<br><br>"NIST - National Voluntary Laboratory Accreditation Program - **NVLAP**" (15) | **OoEaT**:<br>- Laboratory Division<br>> "management of equipment authorization program" (7)<br><br>**NVLAP**:<br>> Electromagnetic Compatibility & Telecommunications (1) | **8** | **OoEaT:**<br>> The two lab and research divisions need double as many employees as the policy section due to complexity (see org chart) -> FTE at Laboratory division: ⅖ * 79 = 32<br>> Within the laboratory division, assume that all responsibilities listed on division site require equal number of FTE -> FTE for "management of equipment authorization program" -> ⅕ * 32 = 7<br><br>**NVLAP**:<br>> 80% of 16 employees working full-time -> 13 FTE<br>> All 19 programmes administered by NIST require equal resources -> 1/19 * 13 = 1 | **0.24** | Consider US generating 29% of global revenue of 33.54 billion in 2023 (Precedence Research 2024) |
| Cybersecurity (Nuclear Energy) | Nuclear Regulatory Commission (2024b) | 17 (second last bullet point9 for audit numbers | Operating Reactors Business Line - **ORBL** (108) | **ORBL**:<br>> Support of Cybersecurity Program (12)<br>> Fitness-for-duty-program (12)<br>> Force-on-force inspection (12) | **36** | **ORBL**: ressources equally used for all duties mentioned -> ⅓ for audit = 36 | **66.67** | 8.6 billion of revenue generated with cyber security applications for the energy sector globally, assume that industrial sector (as mentioned in "end user" category) accounts for 90% of cybersecurity demand (due to interest by hackers) and that demand is distributed |

| Industry | Source | Page (and relevant content on that page) | Relevant programmes and/or units (and FTE where available) | Tasks within programme or unit (and FTE where available) | Total | Potential assumptions (number always rounded up to avoid underestimation) | Total per billion dollars in revenue | Source for revenue (with comments about extraction of numbers) |
|---|---|---|---|---|---|---|---|---|
| | | | | | | | | equally among categories within the industrial category = 0.9 * ⅓ * ⅓ * 8.6 billion = 0.84 billion for cybersecurity for nuclear power plants globally (Allied Market Research 2023)<br><br>Consider that US generates 78.31 billion of 183.10 billion = 42%, therefore estimate market for cybersecurity for nuclear power plants in US at 0.36 billion (Statista 2023b) |
| Cybersecurity (Department of Defense Contractor) | Cyber AB (2024) | Landing page | CMMC AB (19)<br><br>but missing DoD staff still involved, no data | | 19 ?? | All employees listed on LinkedIn and vice versa<br><br>Likely an underestimate given staff at DoD sill working with CMMC AB -> Not included in overview table | 2.34 | Consider US generating 47% of revenue of 17.3 billion globally = 8.13 billion (Coherent Market Insights 2024) |
| Cybersecurity (Power Grids) | Federal Energy Regulatory Commission (2023a) | 56, 57 for overview of tasks, 44 for FTE numbers | **FERC** - Objective 2.2 (254) | **FERC** - Goal 2.2.2, FERC Action 4 (11) | **11** | **FERC**: Resources distributed equally among three goals mentioned under Objective 2.2 and within goal 2.2.2 also distributed equally among four tasks mentioned and that within task 4 of objective 2.2.2 50% go to "lessons learned" and rule-making -> ⅓ * ¼ * ½ * 254 = 11 | **13.58** | 8.6 billion of revenue generated with cyber security applications for the energy sector globally, assume that industrial sector (as mentioned in "end user" category) accounts for 90% of cybersecurity demand (due to interest for hackers) and that demand is distributed equally among categories within the industrial category = 0.9 * ¼ * 8.6 billion = 1.93 billion for |

| Industry | Source | Page (and relevant content on that page) | Relevant programmes and/or units (and FTE where available) | Tasks within programme or unit (and FTE where available) | Total | Potential assumptions (number always rounded up to avoid underestimation) | Total per billion dollars in revenue | Source for revenue (with comments about extraction of numbers) |
|---|---|---|---|---|---|---|---|---|
| | | | | | | | | cybersecurity for bulk power systems (= transmission) globally (Allied Market Research 2023)<br><br>Consider that US generates 78.31 billion of 183.10 billion = 42%, therefore estimate market for cybersecurity for bulk power systems (= transmission) in US at 0.81 billion (Statista 2023b) |
| Large Online Platforms | European Commission (2024) | Landing page, FTE hiring numbers & brief description of responsibility of Directorate F | Directorate of "DG-Connect at EU Commission is responsible (Directorate F) - **DGC**" (NA) | - | **50** | **DGC**:<br>Hiring campaign fully reflects FTE needs for the task<br><br>Not included in final table, given EU-focus | **3.57** | Extracted revenue of platform economies in EU in 2020 = 14 billion (European Council 2024) |

Figure A.21: Top-down estimation of resources and access for each case study

**Top-down estimation: Competence**

| Industry | Source | Page | Share of technical staff | Rationale | Potential assumptions |
|---|---|---|---|---|---|
| Accounting | Securities and Exchange Commission (2022) | - | 0.86 | All roles mentioned on website are technical roles with exception of recent graduate programme -> 1/7 | Jobs equally needed |
| Aviation | Federal Aviation Administration (2021a) | 12 - 17 for share of different staffing categories | 0.86 | Use FTE numbers from FTE analysis above and determine weighted average based on shares of technical and non-technical people per office (see following comment) -> based on description of staffing categories on page 12, define Safety Critical Operational Staff and Safety Technical Specialist Staff as technical and operational support staff as non-technical -> 1370/1822 * 0.85 + 452/1822 * 0.9 = ca. 0.86 | Different staffing categories equally distributed within the sub-division of the individual offices (e.g., same staffing categories in "Certification, inspection, surveillance, investigation, and enforcement activities" as for Flight Standards in general |
| Life Sciences | Food and Drug Administration (2023) | Landing Page | Near 1 | All staffing categories listed on CDHR career page only technical | All staffing categories listed on website |
| Finance | U.S. Securities and Exchange Commission (2023) | 51 | 0.05 | Use FTE numbers from FTE analysis above and determine weighted average based on shares of technical and non-technical people per office (see following comment) -> No roles with finance expertise mentioned in activity description in budget report, some financial expertise might be present in "Legal & Policy Group" -> 0 - 0.05 | - |
| Telco | Federal Communications Commission (2024a)<br><br>National Institute of Standards and Technology (2023)<br><br>LinkedIn (for respective staff listed) | Landing page for task description<br><br>NVLAP section for employee names<br><br>Job title | 0.28 | No official staffing information available -> assume that for Laboratory Division "manage" means less than 20% of technical employees; for NVLAP check LinkedIn profiles of employees listed on website and use study background as proxy for technicality of their current role ->Overall: 0.2 * 7/8 + 1/8 * ⅞ = ca. 0.28 | "Manage" corresponds to technical focus < 20%<br><br>Representation of employees at NVLAP on LinkedIn not correlated with the technicality of their role |
| Cybersecurity (Nuclear Energy) | Nuclear Regulatory Commission (2021b) | Third paragraph on landing page | At least 0.5 | "…focuses recruitment efforts on engineers, scientists and security professionals" | "Focus" corresponds to technical focus < 50% |
| Cybersecurity | No staffing | - | - | - | - |

| Industry | Source | Page | Share of technical staff | Rationale | Potential assumptions |
|---|---|---|---|---|---|
| (Department of Defense Contractor) | information available, estimation in combination with vague FTE estimates would be too uncertain | | | | |
| Cybersecurity (Power Grids) | Federal Energy Regulatory Commission (2022) | Landing page for job profiles | 0.8 | 10 staffing categories mentioned on website, among which 2 are non-technical | Equal hiring of roles |
| Large Online Platforms | European Commission (2024) | Landing page, first paragraph | 0.2 | "40 legal officers, data scientists or technology specialists, and policy and operations specialists, and 10 administrative, policy or legal assistants" will be hired among which data scientists and technology specialists are considered to be technical roles -> 2 out of 4 job categories to be hired within the contingent of 40 people are technical = 0.2 | "Legal officers, data scientists or technology specialists, and policy and operations specialists" to be equally represented among the 40 new-hires<br><br>Hiring efforts fully represent FTE demand for fulfilment of DSA responsibilities |

Figure A.21: Top-down estimation of competence for each case study

## B. Methodological details: Quantification and search protocols

### B.1 Demand-side factors quantification

To enable comparability across industries, the qualitative definitions of the demand-side factors were quantified in two steps. In a first step, a quantitative proxy variable was defined for each demand-side factor (cf. column 2). In a second step, the value range of the proxy variable was divided into three intervals ("standardized value"): high values, medium values, low values (cf. columns 3 - 5) to simplify interpretation and comparability across factors. Proxy-based approach has the advantage that estimation is based on data sources that are accessible to all researchers instead of the actual ones (e.g. verification costs). In future work, a further distinction could be made between observable variables (such as verification costs) and latent variables (such as public salience). While in the first case we already have a high level of confidence due to the established variables used for this purpose, such as the Herfindal index, in the second case it could make sense to use specific methods for analyzing latent variables, such as factor analysis.

| Demand-side Factor | Proxy | Standardized Rating | | |
|---|---|---|---|---|
| | | **High** | **Medium** | **Low** |
| **Scale of risk externality** | Impact and likelihood of risk event | Significant impact, >5% likelihood OR Catastrophic impact, any likelihood | Moderate impact, >5% likelihood OR Significant impact, >1% likelihood | Minor/limited/ moderate impact, <5% likelihood |
| **Verification costs** | Invasiveness of test procedure | Experiments in use case environment | Onsites inspection & experiments in proxy environment **OR** | Simulation of use case environment **AND/OR** outside logic verification |
| | **Source:** Auditing rules | | Onsites inspection & simulation of use case environment & outside logic verification | |
| **Skill specificity** | Annual market-based salary (USD) | >150,000 | 110,000 - 150,000 | 0 - 110,000 |
| **Information sensitivity** | Governmental classification requirements for "product" information | Access restricted to persons with legitimate | Access restricted to persons with legitimate interest within firm | No classification requirements |
| **Risk uncertainty** | ISO Standards (length & share currently under development vis-a-vis existing standards) | >2,000 pages of ISO documentation AND >50% share of ISO standards currently under development | >2,000 pages of ISO documentation OR >50% share of ISO standards currently under development | <2,000 pages of ISO documentation AND <50% ISO standards currently under development |
| **Public salience** | Total Google News search results for 2019-2024 (M) | >15 | 5-15 | <5 |
| **Market concentration** | Herfindahl Index (Points) | >1,000 | 500-1,000 | 0-500 |

Figure B.1: Quantification of demand-side factors

The **decision logic for determining a suitable proxy variable** in step 1 was the following:

1. Does a suitable index already exist within the field of economics (c.f. market concentration)?
2. If not, did a government measure similar factors and publish their results (c.f. scale of risk externality)?
3. If not, did a third-party measure similar factors and publish their results (c.f. skill specificity)?
4. If not, are there government documents that can be used to extract data about the factor and quantify it by defining our own proxy (cf. verification resources, information sensitivity)?
5. If not, are there third-party documents that can be used to extract data about the factor and quantify it by defining our own proxy (cf. risk uncertainty, risk public salience)?

**Scale of risk externality** is assessed by contrasting the likelihood of the risk event with the societal-level impact.

Since we are specifically interested in the risk externality, we infer the risk impact from the UK National Risk Register. As per its mandate, it focuses on the effect of risk incidents on the entire society, making it superior to other proxies, such as liability insurances, which tend to measure the risk internality. **Verification costs** are derived qualitatively from the primary building blocks of the testing procedure and their relative invasiveness. As such, we know that experiments in a true use case environment require substantially more resources to conduct than simulations. Ideally, we would have employed quantitative measures, such as the average cost of an audit in that industry, but unfortunately we could not gain access to such data. **Skill specificity** is inferred from the average private sector salary in a job that requires skills comparable to the typical profile of an auditor in that particular industry. We assume that more specific skills are linked to less labor supply. In turn, economic theory, backed by empirical evidence, predicts that more specific skills, and thus limited labor supply, are associated with higher salaries, at similar levels of labor demand (Broecke, 2016). **Information sensitivity** is assessed by analyzing the qualitative criteria for accessing product information in a given industry. Intuitively, the government prescribes greater access barriers to guard more sensitive information. **Risk uncertainty** is evaluated via the volume of ISO standards, as well as the relative share of standards under development. Generally speaking, higher risk levels should necessitate more standards, intended to manage these risks. At the same time, it seems likely that a certain share of risks remains undetected, thus high risk levels should typically also correspond to somewhat higher risk uncertainty. This is particularly true in very nascent industries, where a substantial share of standards is still under development. **Public salience** is derived from the volume of Google News search results. While the use of internet search data as a proxy for issue salience has its pitfalls, prior research mostly corroborates its robustness (Mellon 2013). As we are primarily interested in the public's attentiveness towards an industry's risks, we limit our search to Google News, thereby excluding Google Search results which for some industries, like aviation, are heavily-driven by consumer offerings, e.g., regarding flights. **Market concentration** is measured via the Herfindahl Index which is among the most commonly applied measures in the economics literature when assessing and comparing industry-level market concentration (Knot & Pasipanodya 2023). Additionally, it is reported at a highly granular-level, down to 5-digit NAICS codes, which allows us to better approximate market concentration for particular applications within the wider industry, which is our analysis' focus.

The **decision logic for determining the proxy variable categories** in step 2 was based on a distribution of the existing case studies, when possible along logical steps or along quartiles. Nevertheless, these categories can be seen as somewhat arbitrary and dependent on the selected case studies.

**We average across demand-side factors** to compare the demand-side factors between the different industries.. To do so, we determined the average across the demand-side factors in two steps building upon our quantification logic. First, we assigned each proxy variable level (high, medium, low) a number. We used "1" for low, "2" for medium and "3" for high. These three chosen numbers have equal distances between each other by design, since "low", "medium" and "high" were defined to have equal distance, too. We then calculated the average across these ordinal values for each case study industry.

### B.2 Auditing responsibilities search procedure

The data extraction procedure for the dimensions of auditing responsibilities and connected history was the following:

1. We searched for the entities involved in auditing of respective product by (A) using Google Search with the search query "Who is responsible for overseeing < product > in the US?" (looked at first page of search results) and (B) ChatGPT with the same query and kept all results that were mentioned by both of them. We never relied solely on ChatGPT.
2. Based on the list of entities, we determined their roles using the definitions from the background section of this appendix and the following sources: a) Information available on the entity website and if not sufficient, b) Federal laws and if not sufficient and c) Third-party information such as newspaper articles.

### B.3 Auditing responsibility logic (Q1)

The decision logic is deducted from the historical case study analysis. Safety considerations come first - often due to public salience - then efficiency considerations in a second step. Demand-side factors referring to Menard (2004) link to effectiveness in terms of safety, other demand-side factors to effectiveness in terms of efficiency.

**Comparison of the degree of regulatory involvement in auditing** across industries was conducted to assess responsibility differences. Industries in which the regulator does all three ("develop", "collect", "judge") - nuclear, aviation, life sciences and big online platforms receive the top 4 ranks. Given the higher involvement of third--parties in collecting information, especially in big online platforms, partly pronounced, in life sciences and also partly in aviation, nuclear is ranked first, then aviation and then life sciences and then big online platforms. Industries in which the regulator does two out of three audit steps - cyber for gov. contractors - are ranked next. In all other industries, the regulator is only involved in developing standards. Thus, we rank them according to the prevalence of the different categories of external parties across the three audit steps. Industries with high third-party involvement (Cyber for power grid) come next. The remaining industries with a mix of first-party and second-party involvement are then Telco, Accounting and Finance. When this quantification is extended to the non-core dimensions of auditing responsibilities described above, the ranking is similar.

### B.4 Resource analysis (Q2) - Bottom-up

We disseminated this worksheet to Advanced AI auditing experts to receive estimates on resources (in terms of #FTE and #weeks) and information access for auditing one Advanced AI model. We received qualitative and quantitative responses from n=11, of which three work at advanced AI Labs, three in teams at regulators doing advanced AI audits, two at academic institutions and three at non-profits doing advanced AI audits. The data revealed that for comprehensive evaluations, especially those involving new threat models or advanced AI, a significant increase in full-time equivalents (FTEs) is required. This varies widely depending on the type of audit. Experts also highlighted the inherent uncertainties in these estimates, primarily due to the evolving nature of AI technologies and the complexity of threat models. Factors such as the readiness of existing infrastructure, the specificity and novelty of the AI capabilities being assessed, and the depth of risk areas all influence the variability in required resources. There is also a pronounced variability in the time needed to set up and interpret evaluations, especially when pioneering new methodologies or dealing with high-risk domains.

### B.5 Resource analysis (Q2) - Top-down

The resources in each industry were determined by following a pre-defined search procedure. Due to data availability restrictions resources, competence and access were determined on an audit level (mixing together "collect" and "judge", while excluding "develop"). First, the search keywords were defined that correspond to our definition of audit: Inspection, Investigation, Oversight, Surveillance, Evaluation, Authorization, Scrutiny.

First, **determine the name of the respective regulatory agency** via results from case studies. Second, **determine the names of the relevant units at the respective regulatory agency** by a) using the following three sources to generate a list of possible units:

1. Google Search: Collect all names mentioned in pages listed on first page of search results for the term "Units at <regulator name> responsible for overseeing <auditee or third-party auditor name, depending on regulatory regime>"
2. Org Chart Analysis: Collect all unit names that have one or more of the search keywords in them.
3. ChatGPT Search: Collect all results given for the prompt "Which units at <regulator name> are responsible for overseeing <auditee or third-party auditor name, depending on regulatory regime>? " All ChatGPT results where confirmed with a web search.

Then:

4. Visit websites on first page of Search results for each of the names mentioned in at least one of the above sources and verify whether audit is mentioned among their scope (see definition of scope at the beginning of this section). If the search results do not encompass units with the desired scope, look into all programmes instead of a subset in the next step.
5. If individual unit websites are not available, keep on the list and verify whether investigations are mentioned among their scope based on information provided in documents consulted in the next step

To e**xtract the number of FTE involved in audit activities at the respective units**

1.Search for congressional workforce planning documents and extract number of FTE working on audit related activities in respective units, as mentioned in report

2. If not available, search for congressional budget reports (or EU equivalent). Extract number of FTE working on audit related activities, as mentioned in report. If number of FTE is not given on the task level, extract the task categories from either the report or the unit's website and evenly divide the number of FTEs by the different tasks (unless it was clear that activities had to be performed in a lab which we assumed to be more resource intensive and therefore assumed that twice the number of FTE was needed)

3. If not available (e.g., due to recent regulatory changes, search for overall number of FTE at agency in charge and conduct individual estimation based on information provided on website (see below for Foundational AI industry and Narrow Hiring AI industry)

To **extract the percentage of technical FTE among the FTE involved in audit activities**:

1.Extract from congressional workforce planning documents.

2. If not available, use information about tasks mentioned in congressional budget report and conduct individual estimation. Technical staff are roles framed as "specialists". "Supportive" roles are non-technical staff. (see industry specific comments in the table)

3. If not available, use information on profiles hired on careers website (and assume that all profiles mentioned are equally needed)

Lastly, to **extract the regulator's information access:** Use values extracted for information access and audit of auditor dimensions of auditing responsibilities

## C. Auditor archetypes

### Auditor Archetype 1: Public Bodies

The public body could conduct some or all auditing directly.

*Independence.* We suggest that the public body as auditor is likely independent. This is because it is neither selected by the auditee nor dependent on payment by the auditee. However, depending on the context, there may be important exceptions including politicization, corruption, and, at the level of individual staff, the incentive of future employment by auditee firms.

*Resources.* The public body might leverage its authority and resources. If the public places importance on the risks posed by the audited industry, the public body's mandate may be emboldened (Ramanna 2015). Conversely, it may be undermined if risks are not publicly salient. Furthermore, we assume a public body's resources cannot scale as rapidly and flexibly as the resources contained within a private auditing market.

*Competence*. Historically, public bodies across industry contexts have developed the competence required to execute their mandates even when competing with the private sector for talent (Lawrence, Cui and Ho 2023). However, they may lack and be unable to quickly access certain kinds of niche expertise, e.g., in new and technically complex auditing methods (Stein and Dunlop 2023).

*Access*. The public body may insist upon a high level of access to evidence controlled by the auditee and required for audits. The public body may serve as a more legitimate and trustworthy custodian of evidence and audit results that are sensitive to national security concerns.

### Auditor Archetype 2: Private Auditors Appointed by Public Bodies

Auditing could be conducted by businesses, civil society, universities and other nongovernmental organizations (hereinafter "private auditors") who are appointed by the public body rather than contracted by auditees.

*Independence.* Under this model, the potential for conflicts of interest between the auditor and auditee are reduced (Fiolleau et al. 2013). However, they are not impossible, for example, if the auditee is the auditor's client for non-audit services (Kowaleski, Mayhew and Tegeler 2018, Raji et al. 2022).

*Resources.* The requirement that auditors be publicly appointed, rather than engaged by auditees, adds a layer of bureaucracy that may cause delays and thereby constrain auditing supply.

*Competence*. Improving on Archetype No. 1, we assume that the public body's access to private auditors taps into and creates incentives for a wider variety of auditing skills and expertise. This may be especially useful where auditing standards and best practices are nascent or evolving (Kurt 2022, Tanner 2000, Galland 2024).

*Access*. Lacking the authority of a public body, private auditors may enjoy less privileged access to the evidence required for auditing. To the extent such evidence is sensitive to national security concerns, private auditors may require security clearances.

### Auditor Archetype 3: Private Auditors Selected by Auditees

Auditing could be both mandatory and left to the market, meaning auditees are free to choose and pay auditors in fulfillment of their auditing obligations. The public body's role in such a model could involve overseeing, setting standards for and accrediting private auditors.

*Independence*. Independence is at risk to the extent the auditor is incentivized by the prospect of repeat business from the auditee (Duflo et al. 2013, Moore, Tanlu and Bazerman 2010, Effing and Hau 2015). A regulatory framework could augment independence by, for example, stipulating that auditors must not have any conflicts of interests with auditees.[8]

*Resources.* We assume an auditing market creates incentives for firms to develop auditing capacity, which can more rapidly scale to meet shifts in auditing demand.

*Competence*. We expect that auditing markets incentivize a wider variety of auditing skills and expertise. In contexts such as life sciences, accreditation systems are utilized to standardize and signal auditor competencies and specializations (CLIA - FDA 2023).

*Access*. Consistent with Archetype No. 2, private auditors may enjoy less privileged access to the evidence required for auditing. Security clearances and systems for managing private sector access to sensitive information are relevant as in Archetype No. 2.

### Auditor Archetype 4: Internal Auditors

Internal auditing refers to auditees evaluating their own systems and technologies. With emerging exceptions, this archetype approximates the status quo in an advanced AI context (Birhane et al. 2024, Weidinger et al. 2023). The question for policymakers is whether a reliance on internal auditing can produce the socially optimal quantity and quality of auditing.

*Independence*. Independence is compromised. Being a part of the auditee, the auditor shares the auditee's pressures to lower costs and minimize barriers to releasing products and services

[8] See, for example, Article 37 of the European Union's Digital Services Act.

(Mutchler 2003). Independence could be augmented by corporate governance and rules stipulating internal auditors be accountable to the board as opposed to management.

*Resources*. We suspect resources to internally audit varies between auditees depending on their size and resources. For example, Google employs close to 200,000[9] employees worldwide and OpenAI employs less than 3,000[10]. Furthermore, if auditees are not obliged to engage external auditors, there may be less incentive for external firms to develop auditing services.

*Competence*. Auditees are in the best position to understand their own systems, especially complex and innovative technologies such as advanced AI. However, they may require external expertise to understand the external impacts of their practices and products, for example, on elections and citizens' health and safety.

*Access*. Auditees are often in the best position to access the evidence required for audits, which is in their control. This presumes that the access of internal auditors is not restricted by gatekeepers within the auditee.

[9] See "How Many Employees Does Google Have?" Doofinder, https://www.doofinder.com/en/statistics/how-may-employees-does-google-have.

[10] See "OpenAI Employee Directory, Headcount & Staff" LeadIQ, https://shorturl.at/2Q1E8.

## References

Alaghbari, K. A.; Saad, M. H. M.; Hussain, A.; and Alam,

M. R. 2022. Complex event processing for physical and cyber security in datacentres - recent progress, challenges and recommendations. Journal of Cloud Computing, 11(1): 65.

Allied Market Research, A. M. R. 2023. Cyber Security in Energy Sector Market Size, Forecast - 2032.

Anderljung, M.; Smith, E. T.; O'Brien, J.; Soder, L.; Bucknall, B.; Bluemke, E.; Schuett, J.; Trager, R.; Strahm, L.; and Chowdhury, R. 2023. Towards Publicly Accountable Frontier LLMs: Building an External Scrutiny Ecosystem under the ASPIRE Framework. ArXiv:2311.14711 [cs].

Anthropic. 2023. Challenges in evaluating AI systems \ Anthropic.

Anthropic. 2024. Model Card Claude 3.pdf.

Ayres, I.; and Braithwaite, J. 1992. Responsive Regulation: Transcending the Deregulation Debate. New York: Oxford University Press. Print.

Barrett, Anthony M.; Newman, Jessica; Nonnecke, Brandie; Hendrycks, Dan; Murphy, Evan R.; and Jackson, Krystal. 2023. Berkeley-GPAIS-Foundation-Model-Risk- Management-Standards-Profile-v1.0.pdf.

Behr, R. L.; and Iyengar, S. 1985. Television News, Real- World Cues, and Changes in the Public Agenda. Public Opinion Quarterly, 49(1): 38–57.

Birhane, A.; Steed, R.; Ojewale, V.; Vecchione, B.; and Raji, I. D. 2024. AI auditing: The Broken Bus on the Road to AI Accountability. ArXiv:2401.14462 [cs].

Blind, K. 2006. Explanatory factors for participation in formal standardisation processes: Empirical evidence at firm level. Economics of Innovation and New Tech- nology, 15(2): 157–170. Publisher: Routledge eprint: https://doi.org/10.1080/10438590500143970.

Blind, K. 2013. The Impact of Standardization and Standards on Innovation.

Bos, G. 2018. ISO 13485:2003/2016—Medical Devices—Quality Management Systems—Requirements for Regulatory Purposes. In Handbook of Medical Device Regulatory Affairs in Asia. Jenny Stanford Publishing, 2 edition. ISBN 978-0-429-50439-6. Num Pages: 22.

Bourgeois, L. J. 1981. On the Measurement of Organizational Slack. The Academy of Management Review 6(1): 29–39. doi.org/10.2307/257138. Accessed: 2024-07-25.

OECD 2014. Measuring Environmental Policy Stringency in OECD Countries: A Composite Index Approach. Technical report, OECD, Paris.

Broecke, S. 2016. Do skills matter for wage inequality? IZA World of Labor.

Brundage, M.; Avin, S.; Wang, J.; Belfield, H.; Krueger, G.; Hadfield, G.; Khlaaf, H.; Yang, J.; Toner, H.; Fong, R.; Maharaj, T.; Koh, P. W.; Hooker, S.; Leung, J.; Trask, A.; Bluemke, E.; Lebensold, J.; O'Keefe, C.; Koren, M.; Ryffel, T.; Rubinovitz, J. B.; Besiroglu, T.; Carugati, F.; Clark, J.; Eckersley, P.; de Haas, S.; Johnson, M.; Laurie, B.; Ingerman, A.; Krawczuk, I.; Askell, A.; Cammarota, R.; Lohn, A.; Krueger, D.; Stix, C.; Henderson, P.; Graham, L.; Prunkl, C.; Martin, B.; Seger, E.; Zilberman, N.; hE´ igeartaigh, S. ; Kroeger, F.; Sastry, G.; Kagan, R.; Weller, A.; Tse, B.; Barnes, E.; Dafoe, A.; Scharre, P.; Herbert-Voss, A.; Rasser, M.; Sodhani, S.; Flynn, C.; Gilbert, T. K.; Dyer, L.; Khan, S.; Bengio, Y.; and Anderljung, M. 2020. Toward Trustworthy AI Development: Mechanisms for Supporting Verifiable Claims. ArXiv:2004.07213 [cs].

Casper, S.; Ezell, C.; Siegmann, C.; Kolt, N.; Curtis, T. L.; Bucknall, B.; Haupt, A.; Wei, K.; Scheurer, J.; Hobb- hahn, M.; Sharkey, L.; Krishna, S.; Von Hagen, M.; Al- berti, S.; Chan, A.; Sun, Q.; Gerovitch, M.; Bau, D.; Tegmark, M.; Krueger, D.; and Hadfield-Menell, D. 2024. Black-Box Access is Insufficient for Rigorous AI Audits. ArXiv:2401.14446 [cs].

Coase, R. H. 1960. The Problem of Social Cost. The Journal of Law & Economics, 3: 1–44. Publisher: [University of Chicago Press, Booth School of Business, University of Chicago, University of Chicago Law School].

Code of Federal Regulations Title 14, U. S. 2024. 14 CFR Part 21 – Certification Procedures for Products and Articles.

Code of Federal Regulations Title 16, U. S. 2024. eCFR :: 16 CFR Part 314 – Standards for Safeguarding Customer Information.

Code of Federal Regulations Title 47, U. S. 2024. 47 CFR § 68.162 Requirements for Telecommunication Certification Bodies.

Coherent Market Insights, U. S. 2024. Defense Cyber Security Market - Price, Size, Share & Growth.

Cohen, M. K.; et al. 2024. Regulating Advanced Artificial Agents. Science 384: 36–38. doi.org/10.1126/science.adl0625.

Costanza-Chock, S.; Harvey, E.; Raji, I. D.; Czernuszenko, M.; and Buolamwini, J. 2023. Who Audits the Auditors? Recommendations from a field scan of the algorithmic auditing ecosystem. In 2022 ACM Conference on Fairness, Accountability, and Transparency, 1571–1583. ArXiv:2310.02521 [cs].

Crawford. 2022. Atlas of AI.

Cyber AB, U. S. 2024. The Cyber AB: Overview.

Davenport, C. 2023. SpaceX to the FAA: The industry needs you to move faster. Washington Post.

DeFond, M. L. 2010. How should the auditors be audited? Comparing the PCAOB Inspections with the AICPA Peer Reviews. Journal of Accounting and Economics, 49(1): 104–108.

DeFond, M.; and Zhang, J. 2014. A Review of Archival Auditing Research. Journal of Accounting and Economics 58(2–3): 275–326. doi.org/10.1016/j.jacceco.2014.09.002.

Dennis, K. 2008. The Rating Game: Explaining Rating Agency Failures in the Buildup to the Financial Crisis. University of Miami Law Review, 63(4): 1111–1150.

Department of Defense, U. S. 2024. About CMMC.

Department of Science, Innovation and Technology (DSIT), U.K Government. 2024. International Scientific Report on the Safety of Advanced AI.

Digital Services Act, E. 2022.

Duflo, E.; Greenstone, M.; Pande, R.; and Ryan, N. 2013. Truth-telling by Third-party Auditors and the Response of Polluting Firms: Experimental Evidence from India*. The Quarterly

Journal of Economics, 128(4): 1499–1545. Publisher: Oxford Academic.

Efing, M.; and Hau, H. 2015. Structured debt ratings: Evidence on conflicts of interest. Journal of Financial Economics, 116(1): 46–60.

EpochAI, E. 2023. Machine Learning Trends.

Bol, A., Grabner, I., Vienna, W., & Haesebrouck, K. 2019. Literature review The effect of audit culture on audit quality. Found Audit Res.

EU. 2022. The Digital Services Act (DSA) - Regulation (EU) 2022/2065.

European Commission, E. 2018. DocsRoom European Commission.

European Commission, E. 2024. Do you want to help enforce the Digital Services Act? Apply now to be part of the DSA enforcement team! | Shaping Europe's digital future.

European Council, E. 2024. The EU's platform economy.

European Court of Auditors, E. 2015. EU supervision of credit rating agencies – well established but not yet fully effective. credit rating agencies.

European Union Aviation Safety Agency, E. 2024. Aircraft certification | EASA.

FDA 2023. CLIA Categorizations. FDA. Publisher: FDA.

Federal Aviation Administration, U. S. 2021a. Aviation Safety Workforce Plan 2021 | 2030.

Federal Aviation Administration, U. S. 2021b. A Brief History of the FAA | Federal Aviation Administration.

Federal Aviation Administration, U. S. 2021c. Legal Enforcement Actions | Federal Aviation Administration.

Federal Aviation Administration, U. S. 2022. Airworthiness Certification Overview | Federal Aviation Administration.

Federal Aviation Administration, U.S. 2022. Bilateral Agreements | Federal Aviation Administration.

Federal Communications Commission, U. S. 2015. Laboratory Division | Federal Communications Commission.

Federal Communications Commission (webpage). Equipment Authorization – RF Device | Federal Communications Commission.

Federal Communications Commission (webpage). Equipment Authorization Procedures | Federal Communications Commission.

Federal Communications Commission, U. S. 2023. Office of Engineering and Technology (OET) Organization Chart

| Federal Communications Commission.

Federal Communications Commission, U. S. 2024a. DOC-391605A1.pdf.

Federal Communications Commission, U. S. 2024b. Enforcement Primer | Federal Communications Commission.

Federal Communications Commission, U. S. 2024c. Equipment Authorization | Federal Communications Commission.

Federal Communications Commission, U. S. 2024d. Main webpage (William H. Donaldson).

Federal Energy Regulatory Commission, U. S. 2022. Career Opportunities | Federal Energy Regulatory Commission.

Federal Energy Regulatory Commission, U. S. 2023a. FERC FY 24 Congressional Justification | Federal Energy Regulatory Commission.

Federal Energy Regulatory Commission, U. S. 2023b. Re- lated Document Classes | Federal Energy Regulatory Commission.

Federal Energy Regulatory Commission, U. S. 2023c. Reliability Explainer | Federal Energy Regulatory Commission.

Financial Industry Regulatory Authority, U. S. 2010. 4140. Audit | FINRA.org.

Financial Industry Regulatory Authority, U. S. 2024. What We Do | FINRA.org.

Fiolleau, K.; Hoang, K.; Jamal, K.; and Sunder, S. 2013. How Do Regulatory Reforms to Enhance Auditor Independence Work in Practice? Contemporary Accounting Research, 30(3): 864–890. eprint:
https://onlinelibrary.wiley.com/doi/pdf/10.1111/1911-3846.12004.

Food and Drug Administration, U. S. 2018. Details of Full- Time Equivalents.

Food and Drug Administration, U. S. 2023. Jobs at the Center for Devices and Radiological Health (CDRH). Publisher: FDA.

Food and Drug Administration, U. S. 2024a. 510(k) Third Party Review Program. Publisher: FDA.

Food and Drug Administration, U. S. 2024b. CDRH Management Directory by Organization. FDA. Publisher: FDA.

Food and Drug Administration, U. S. 2024c. FDA's Risk- Based Approach to Inspections. FDA. Publisher: FDA.

Food and Drug Administration, U. S. 2024d. A History of Medical Device Regulation & Oversight in the United States. FDA. Publisher: FDA.

Food and Drug Administration, U. S. 2024e. Overview of Device Regulation. Publisher: FDA.

Fortune Business Insights, U. S. 2024. U.S. Medical Devices Market Size, Share | Analysis Report, 2030.

Galland, J.-P. 2024. Standards, Certification, and Accreditation: Indispensable Tools for European Safety Regulations? In The Regulator–Regulatee Relationship in High-Hazard Industry Sectors, 71–78. Springer, Cham. ISBN 978-3-031- 49570-0. ISSN: 2191-5318.

Glassdoor. 2024. Company Salaries.

Government Accountability Office, U. S. 2021. Securities Regulation: SEC Could Take Further Actions to Help Achieve Its FINRA Oversight Goals | U.S. GAO.

Groves, L.; Metcalf, J.; Kennedy, A.; Vecchione, B.; and Strait, A. 2024. Auditing Work: Exploring the New York City algorithmic bias audit regime. ArXiv:2402.08101 [cs].

Gunningham, N.; Grabosky, P.; and Sinclair, D. 1998. Smart Regulation: Designing Environmental Policy. Oxford: Oxford Academic. doi.org/10.1093/oso/9780198268574.001.0001.

Hadfield, G.; Clark, J. 2023. Regulatory Markets: The Future of AI Governance. ArXiv:/2304.04914 [cs]

Hansen, S. C.; Kumar, K. R.; and Sullivan, M. W. 2008. Auditor Capacity Stress and Audit Quality: Market-Based Evidence from Andersen's Indictment.

Hazlett, T.; and Pai, A. 2018. The Untold History of FCC Regulation.

Hill, M.; and Varone, F. 2021. The Public Policy Process. London: Routledge, 8 edition. ISBN 978-1-00-301020-3.

Hobbhahn, M.; and Scheurer, J. 2024. Apollo Research. We need a Science of Evals.

Huang, K.; Wang, Y.; Goertzel, B.; Li, Y.; Wright, S.; and Ponnapalli, J., eds. 2024. Generative AI Security: Theories and Practices. Future of Business and Finance. Cham: Springer Nature Switzerland. ISBN 978-3-031-54251-0 978-3-031-54252-7.

IEEE. 2008. IEEE Standard for Software Reviews and Audits. ISBN: 9780738157689.

ISO/IEC. 2015. ISO/IEC 17021-1:2015 - Conformity assessment - Requirements for bodies providing audit and certification of management systems - Part 1: Requirements.

ISO/IEC. 2022. ISO/IEC 27001:2022 - Information security, cybersecurity and privacy protection — Information security management systems — Requirements.

Jorgenson, D. W.; Landefeld, J. S.; and Schreyer, P. 2014. Measuring Economic Sustainability and Progress. Univer- sity of Chicago Press. ISBN 978-0-226-12133-8 978-0-226- 12147-5.

Kleinman, G.; Lin, B. B.; and Palmon, D. 2014. Audit Quality: A Cross-National Comparison of Audit Regulatory Regimes. Journal of Accounting, Auditing & Finance, 29(1): 61–87. Publisher: SAGE Publications Inc.

Kowaleski, Z. T.; Mayhew, B. W.; and Tegeler, A. C. 2018. The Impact of Consulting Services on Audit Quality: An Experimental Approach. Journal of Accounting Research, 56(2): 673–711. eprint: https://onlinelibrary.wiley.com/doi/pdf/10.1111/1475-679X.12197.

Kurt, A. C. 2022. Audit Risk and the Implications of Employing Specialist Auditors: Evidence from Government Contractors.

Lamoreaux, P. T. 2016. Does PCAOB inspection access improve audit quality? An examination of foreign firms listed in the United States. Journal of Accounting and Economics, 61(2): 313–337.

Lawrence, C.; Cui, I.; and Ho, D. 2023. The Bureaucratic Challenge to AI Governance: An Empirical Assessment of Implementation at U.S. Federal Agencies. In Proceedings of the 2023 AAAI/ACM Conference on AI, Ethics, and Society, 606–652. Montr\'{e}al QC Canada: ACM. ISBN 9798400702310.

Lennon, H. 2021. Why The SEC's Stance On Bitcoin ETFs May Need To Change. Section: Crypto & Blockchain.

Levi-Faur, D. 2003. Comparative Research Designs in the Study of Regulation: How to Increase the Number of Cases Without Compromising the Strengths of Case- Oriented Analysis.

Levi-Faur, D. 2011. Regulation and Regulatory Governance. In Handbook on the Politics of Regulation, edited by D. Levi-Faur, Chapter 1. Cheltenham, UK: Edward Elgar Publishing.

Liang, P.; Bommasani, R.; Lee, T.; Tsipras, D.; Soylu, D.; Yasunaga, M.; Zhang, Y.; Narayanan, D.; Wu, Y.; Kumar, A.; Newman, B.; Yuan, B.; Yan, B.; Zhang, C.; Cosgrove, C.; Manning, C. D.; Re´, C.; Acosta-Navas, D.; Hudson, D. A.; Zelikman, E.; Durmus, E.; Ladhak, F.; Rong, F.; Ren, H.; Yao, H.; Wang, J.; Santhanam, K.; Orr, L.; Zheng, L.; Yuksekgonul, M.; Suzgun, M.; Kim, N.; Guha, N.; Chatterji, N.; Khattab, O.; Henderson, P.; Huang, Q.; Chi, R.; Xie, S. M.; Santurkar, S.; Ganguli, S.; Hashimoto, T.; Icard, T.; Zhang, T.; Chaudhary, V.; Wang, W.; Li, X.; Mai, Y.; Zhang, Y.; and Koreeda, Y. 2023. Holistic Evaluation of Language Models. ArXiv:2211.09110 [cs].

Marco, A. C.; Carley, M.; Jackson, S.; and Myers, A. 2017. The USPTO Historical Patent Data Files: Two Centuries of Innovation.

Maynard, M. 2014. The FCC TCB Program: A Government and Industry Cooperative.

MDC. 2022. Price List (Certification according to MDR).

Mellon, J. 2013. Internet Search Data and Issue Salience: The Properties of Google Trends as a Measure of Issue Salience. Journal of Elections, Public Opinion and Parties, 24(1): 45–72. Publisher: Routledge eprint: https://doi.org/10.1080/17457289.2013.846346.

METR. 2024. Portable Evaluation Tasks via the METR Task Standard.

Moore, D.; Tanlu, L.; and Bazerman, M. 2010. Conflict of Interest and the Intrusion of Bias. Judgment and Decision Making, 5: 37–53.

Mutchler, J. F. 2003. Chapter 7: Independence and Objectivity - A Framework for Research Opportunities in Internal Auditing.

Menard, C. 2004. The Economics of Hybrid Organizations. Journal of Institutional and Theoretical Economics (JITE) / Zeitschrift fu¨r die gesamte Staatswissenschaft, 160(3): 345–376. Publisher: Mohr Siebeck GmbH & Co. KG.

Moekander, J.; Schuett, J.; Kirk, H. R.; and Floridi, L. 2023. Auditing large language models: a three-layered approach. AI and Ethics.

National Institute of Standards and Technology, U. S. 2023. Staff. NIST.

National Institute of Standards and Technology, U. S. 2024a. Accreditation Programs. NIST. Last Modified: 2023-05-17T10:59-04:00.

National Institute of Standards and Technology, U. S. 2024b. Government Contractor Requirements. NIST. Last Modified: 2024-02-07T10:09-05:00.

National Institute of Standards and Technology, U. S. 2024c. Telecommunications Certification Bodies (TCB) Application Information. NIST. Last Modified: 2021-06-02T18:27- 04:00.

National Risk Register, U. K. 2023. 2023 NATIONAL RISK REGISTER NRR.pdf.

National Science and Technology Council, U.S. Government. 2024. Critical and Emerging Technology List Update.

Neuman, J. 2008. FAA's 'culture of coziness' targeted in airline safety hearing. Section: Travel & Experiences.

Nevo, S.; Lahav, D.; Karpur, A.; Alstott, J.; and Matheny,

J. 2023. Securing Artificial Intelligence Model Weights: Interim Report. RAND Corporation.

Nextlabs.2016. SB-NERC-and-FERC-Cyber-Security.pdf.

Noll, R. G. 1989. Max D. Paglin, ed., A legislative history of the communications Act of 1934: (Oxford University Press, New York, 1989) pp. xxii+981. Information Economics and Policy, 4(2): 190–194.

North American Electric Reliability Corporation, U. S. 2023. Reliability Principles.pdf.

NQA. 2024a. AS Certifications - AS9100 / AS9110 / AS9120 / AS6081 | NQA.

NQA. 2024b. AS9100 Certification - Aerospace Management Standard | NQA.

Nuclear Regulatory Commission, U. S. 2021a. Information Security.

Nuclear Regulatory Commission, U. S. 2021b. Opportunities.

Nuclear Regulatory Commission, U. S. 2021c. § 73.54 Protection of digital computer and communication systems and networks.

Nuclear Regulatory Commission, U. S. 2024a. Cybersecurity | NRC.gov.

Nuclear Regulatory Commission, U. S. 2024b. ML24061A093.pdf.

Ojewale, V.; Steed, R.; Vecchione, B.; Birhane, A.; and Raji, I. D. 2024. Towards AI Accountability Infrastructure: Gaps and Opportunities in AI Audit Tooling. ArXiv:2402.17861 [cs].

OpenAI. 2023. gpt-4-system-card.pdf.

Pasztor. 2023. Opinion: The FAA's Safety System Is Starting to Show Cracks | Aviation Week Network.

Pawlson, L. G.; Torda, P.; Roski, J.; and O'Kane, M. E. 2005. The role of accreditation in an era of market-driven accountability. The American Journal of Managed Care, 11(5): 290–293.

PCAOB. 2024. Standards | PCAOB.

Petrick, J. A.; and Scherer, R. F. 2003. The Enron Scandal and the Neglect of Management Integrity Capacity. American Journal of Business, 18(1): 37–50. Publisher: MCB UP Ltd.

Pigou. 1920. The Economics of Welfare.

Politico. 2024. Rishi Sunak promised to make AI safe. Big Tech's not playing ball.

Power, M. 1999. The Audit Society: Rituals of Verification. Oxford University Press. ISBN 978-0-19-168518-7.

Quélin, B. V.; Cabral, S.; Lazzarini, S.; and Kivleniece, I. 2019. The Private Scope in Public–Private Collaborations: An Institutional and Capability-Based Perspective. Organization Science, 30(4): 831–846. Publisher: INFORMS.

Radu, R. 2021. Steering the governance of artificial intelligence: national strategies in perspective. Policy and Society, 40(2): 178–193.

Rajala, T.; and Kokko, P. 2021. Biased by design – the case of horizontal accountability in a hybrid organization. Accounting, Auditing & Accountability Journal, 35(3): 830– 862. Publisher: Emerald Publishing Limited.

Raji, I. D.; Smart, A.; White, R. N.; Mitchell, M.; Gebru, T.; Hutchinson, B.; Smith-Loud, J.; Theron, D.; and Barnes, P. 2020. Closing the AI accountability gap: defining an endto-end framework for internal algorithmic auditing. In Proceedings of the 2020 Conference on Fairness, Accountability, and Transparency, 33–44. Barcelona Spain: ACM. ISBN 978-1-4503-6936-7.

Raji, I. D.; Xu, P.; Honigsberg, C.; and Ho, D. E. 2022. Outsider Oversight: Designing a Third Party Audit Ecosystem for AI Governance. ArXiv:2206.04737 [cs].

Ramanna, K. 2015. Thin Political Markets: The Soft Underbelly of Capitalism. California Management Review, 57(2): 5–19. Publisher: SAGE Publications Inc.

ReedSmith. 2024. Hiring Non-U.S. Citizens – Don't Forget to Get Your Export Licenses | Perspectives | Reed Smith LLP.

Precedence Research. 2023. Aerospace Market Size To Reach USD 678.17 Billion By 2032.

Precedence Research 2024. RF Components Market Size To Hit USD 101.09 Bn By 2032.

Princeton UniversityUniversity, P. 2024. Sharing Technical Information or Software.

Ryan, J. 2012. The Negative Impact of Credit Rating Agencies and proposals for better regulation.

Sastry, G.; Heim, L.; Belfield, H.; Anderljung, M.; Brundage, M.; Hazell, J.; O'Keefe, C.; Hadfield, G. K.; Ngo, R.; Pilz, K.; Gor, G.; Bluemke, E.; Shoker, S.; Egan, J.; Trager, R. F.; Avin, S.; Weller, A.; Bengio, Y.; and Coyle, D. 2024. Computing Power and the Governance of Artificial Intelligence. ArXiv:2402.08797 [cs].

S&P Global RatingsRatings, S. G. 2022. guide to credit rating essentials digital.pdf.

Schelker, M. 2010. Auditor Expertise: Evidence from the Public Sector.

Schuett, J. 2023. Three lines of defense against risks from AI. AI & SOCIETY. ArXiv:2212.08364 [cs].

Securities and Exchange Commission, U. S. 2020a. EJR Ex 2 Methodologies.pdf.

Securities and Exchange Commission, U. S. 2020b. SEC.gov | The SEC's Office of Credit Ratings and NRSRO Regulation: Past, Present, and Future.

Securities and Exchange Commission, U. S. 2022. SEC.gov | Types of Appointment Authorities.

Securities and Exchange Commission, U. S. 2023. fy-2024-congressional-budget-justification final-3-10.pdf.

Short, J. L.; Toffel, M. W.; and Hugill, A. R. 2016. Monitoring global supply chains. Strategic Management Journal, 37(9): 1878–1897. eprint: https://onlinelibrary.wiley.com/doi/pdf/10.1002/smj.2417.

Simnett, R.; Carson, E.; and Vanstraelen, A. 2016. International Archival Auditing and Assurance Research: Trends, Methodological Issues, and Opportunities. AUDITING: A Journal of Practice & Theory, 35(3): 1–32.

Simnett, R.; and Trotman, K. T. 2018. Twenty-Five-Year Overview of Experimental Auditing Research: Trends and Links to Audit Quality. Behavioral Research in Accounting, 30(2): 55–76.

Solomon, S. D. 2010. The Government's Elite and Regulatory Capture. Section: Business Day.

Statista. 2023a. Cybersecurity - United States | Statista Market Forecast.

Statista. 2023b. Topic: Accounting industry in the U.S.

Statista. 2024a. Forecast: Industry revenue of "securities brokerage" in the U.S. 2012-2024.

Statista. 2024b. Generative AI - North America | Statista Market Forecast.

Stein, M.; and Dunlop, C. 2023. Safe before sale.

Stein, M.; and Dunlop, C. 2024. Safe beyond sale: Post-deployment monitoring of AI.

Stigler, G. J. 1971. The Theory of Economic Regulation. The Bell Journal of Economics and Management Science 2(1): 3–21. doi.org/10.2307/3003160. Accessed: 2024-07-25.

Talent. 2024. Salary in USA - Average Salary.

Tanner, B. 2000. Independent assessment by third-party certification bodies. Food Control, 11(5): 415–417.

The White House, U. S. 2009. Executive Order 13526. Classified National Security Information.

The White House, U. S. 2023. Executive Order on the Safe, Secure, and Trustworthy Development and Use of Artificial Intelligence.

Tortoise. 2023. The Global AI Index - Tortoise.

UK Government. 2023. Safety and security risks of generative artificial intelligence to 2025 (Annex B).

Van Loo, C. L. 2019. Regulatory Monitors: Policing Firms in the Compliance Era.

Watkins, L. . 2022. SEC Investigations.

Weidinger, L.; Rauh, M.; Marchal, N.; Manzini, A.; Hendricks, L. A.; Mateos-Garcia, J.; Bergman, S.; Kay, J.; Griffin, C.; Bariach, B.; Gabriel, I.; Rieser, V.; and Isaac, W. 2023. Sociotechnical Safety Evaluation of Generative AI Systems. ArXiv:2310.11986 [cs].

Weidinger, L.; Uesato, J.; Rauh, M.; Griffin, C.; Huang, P.S.; Mellor, J.; Glaese, A.; Cheng, M.; Balle, B.; Kasirzadeh, A.; Biles, C.; Brown, S.; Kenton, Z.; Hawkins, W.; Stepleton, T.; Birhane, A.; Hendricks, L. A.; Rimell, L.; Isaac, W.; Haas, J.; Legassick, S.; Irving, G.; and Gabriel, I. 2022. Taxonomy of Risks posed by Language Models. In Proceedings of the 2022 ACM Conference on Fairness, Accountability, and Transparency, FAccT '22, 214–229. New York, NY, USA: Association for Computing Machinery. ISBN 978-1- 4503-9352-2.

White, L. J. 2018. The Credit Rating Agencies and Their Role in the Financial System. Book Title: The Oxford Handbook of Institutions of International Economic Governance and Market Regulation Edition: 1 ISBN: 9780190900571 9780190900601 Publisher: Oxford University Press